\documentclass[preprint,12pt,authoryear]{elsarticle}

\usepackage{amssymb}
\usepackage{amsmath}
\usepackage{commath}
\usepackage{siunitx}
\PassOptionsToPackage{hyphens}{url}\usepackage{hyperref}
\usepackage{natbib}
\usepackage{longtable}
\usepackage{booktabs}
\usepackage{makecell}
\usepackage[right]{lineno}
\usepackage{changes}
\definechangesauthor[name={Author}, color=blue]{YL}

\journal{Ocean Modelling}

\begin{document}

\begin{frontmatter}



\title{Machine Learning Based Mesh Movement for
Non-Hydrostatic Tsunami Simulation} 


\author{Yezhang Li\corref{cor1}}
\ead{yezhang.li22@imperial.ac.uk}
\cortext[cor1]{Corresponding author}
\author{Stephan C. Kramer}
\author{Matthew D. Piggott}

\affiliation{organization={Department of Earth Science and Engineering, Imperial College London},
            addressline={South Kensington Campus}, 
            city={London},
            postcode={SW7 2AZ},
            country={United Kingdom}}

\begin{abstract}
This study investigates the use of machine learning based mesh movement method, specifically the Universal Mesh Movement Network (UM2N), with depth integrated non-hydrostatic shallow water models. Motivation for this comes from the need for models which balance efficiency and accuracy for use in probabilistic coastal hazard assessment. Implementations are built on the discontinuous Galerkin finite-element (DG-FE) based software, Thetis, which leverages the  partial differential equation (PDE) framework Firedrake for automated code generation. Verification on benchmark test cases and validation against laboratory measurements of coastal hazards, focusing on tsunami propagation, run-up, and inundation is performed. In these tests, the UM2N-driven meshes help resolve key non-hydrostatic dynamics including wave refraction over a conical shoal, run-up with wetting-drying on a conical island, and tsunami inundation in the Monai Valley laboratory benchmark, and yield numerical solutions in close agreement with reference fine-mesh computations and measured data. Notably, in the Monai Valley case, UM2N achieves a $\sim91\,\%$ reduction in wave-peak error at the nearshore gauge compared with $\sim74\,\%$ for the conventional Monge--Amp\`{e}re (MA) mesh movement, both relative to the coarse fixed mesh. The UM2N surrogate based approach accelerates the conventional mesh movement step, achieving a $\sim32\,\%$ reduction in total runtime and $\sim2\times$ speed-up in mesh movement step time over the MA solver on GPU, while offering a significant improvement in robustness over long integration periods and under strongly nonlinear wave conditions.
\end{abstract}




\end{frontmatter}
\section{Introduction}
\label{sec1}
Accurate modeling of coastal tsunami has been a challenging research topic for many decades, with a goal of effective hazard assessments in the face of complex coastal topography, changeable multi-scale flow dynamics and the impacts of flooding and drying processes. This has been demonstrated by recent events such as the January 2024 Noto Peninsula (Japan) tsunami~\citep{Heidarzadeh2024}, as well as by deterministic hazard assessment efforts for vulnerable coastal regions such as the study of \cite{Ren2022}. Unlike hydrostatic models that are commonly used for large-scale tsunami simulations, it is necessary to consider non-hydrostatic effects in coastal tsunami scenarios in order to represent with sufficient accuracy complex nearshore wave processes, especially for both short waves and relatively long waves with high non-linearity. According to \cite{PAN201968}, the propagation of strong dispersive or non-linear waves towards the shore may undergo significant wave steepening and transformation due to variations in water depth or bottom friction. The associated shoaling and funnelling primarily arise from the combined effects of frequency dispersion caused by non-hydrostatic impacts and wave-wave or wave-structure interactions from non-linearity \citep{CUI201255, PAN201968}. However, it is difficult to represent these phenomena in non-linear shallow water equation (NSWE) models under hydrostatic assumptions. Hence, it is important to consider non-hydrostatic effects in coastal tsunami simulation. A number of non-hydrostatic coastal ocean models have been developed via adopting the approach introduced by \cite{Zijlema2003}. In addition, several Boussinesq wave models have been developed to enhance their applicability in capturing wave diffraction and refraction for coastal wave propagation (See \cite{WeiKirby1995}, \cite{Liang2013} and \cite{shi2013funwave}). This is because the original Boussinesq-type equations take weak dispersion into consideration.

However, the increased accuracy of non-hydrostatic modeling comes with a substantial computational cost, impacting on feasible mesh resolutions, simulation time scales and domain sizes, and the number of simulations that can be conducted. Scientists and engineers in computational fluid dynamics have always sought a balance between accuracy and computational costs in their models. In coastal ocean modelling, a series of simplified non-hydrostatic models have been developed to reduce computational costs, instead of solving the full 3D Navier-Stokes (N-S) equations. According to \cite{PAN201968} and \cite{Stansby1998}, one of the popular simplified approaches is a type of non-hydrostatic model that decomposes the total pressure into a hydrostatic part and a non-hydrostatic part, in which the non-hydrostatic pressure term is retained in the hydrostatic equations to capture the effect of vertical flow acceleration. In Thetis, there are also DG-FE based non-hydrostatic modelling approaches via both multi-layer and depth-integrated systems developed by \cite{PAN201968} based upon the pioneering work of DG-FE based approaches (following \cite{Thetis}) and layer-integrated non-hydrostatic models (see \cite{Zijlema2003}, \cite{WeiJia} and \cite{CUI201255}) for coastal ocean problems. While such non-hydrostatic approaches still have  limitations in capturing wave breaking/overturning, their favourable properties can be quite significant for large-scale applications in coastal ocean/tsunami simulation.

On the other hand, although significant progress has been made in the development of non-hydrostatic models for coastal ocean simulations, most of them rely on fixed and quasi-uniform meshes such as SWASH \citep{ZIJLEMA2011992} and NHWAVE \citep{MA201222}. The main limitation of structured/uniform meshes is the relatively poor representation of the complex coastline and coastal topography, and inflexibility to focus resolution and dynamically adaptation on time-varying changes in the fluid dynamics, especially when tracking wave propagation (usually tsunami) across the ocean \citep{PIGGOTT200595, LeVeque2016}. An alternative way to resolve this problem is to adopt mesh adaptation methods, which allows for mesh resolution to be optimized in order to maximize accuracy while minimizing computational costs, and allowing for large geographical scale simulations to be conducted at the same time as representing detailed coastal processes. In general, there are two main types of mesh adaptation methods: h-adaptation and r-adaptation. Broadly speaking, classical h-adaptive methods refine the mesh resolution by locally adding or removing mesh nodes, which changes the local topological structure. This approach offers great flexibility in concentrating resolution where needed, but the changing connectivity requires interpolation between old and new meshes, which may introduce low solution quality and reduce overall efficiency~\citep{PIGGOTT200595}. While r-adaptive methods change mesh resolution by moving the locations of mesh nodes but fixing the total number of nodes and their connectivity. Since the mesh topology remains unchanged, r-adaptive methods avoid the overhead of re-meshing and are suited to time-dependent problems where the solution features evolve continuously, although the total resolution is limited by the fixed node count~\citep{PIGGOTT200595, Budd2009, Andrew2018}. A detailed comparison of h, r, and combined hr-adaptivity strategies for ocean modelling applications can be found in \cite{PIGGOTT200595}. In the past decade, according to \cite{Wallwork2021}, the use of optimal transport mesh movement driven by the solutions of Monge–Amp\`{e}re (MA) type equations, related to r-adaptivity, has been advocated in the simulation of geophysical flows by \cite{Budd2009} and \cite{Andrew2018}, etc. This enables mesh movement methods based on the MA equations which can also be applied in solving tsunami propagation problems based on NSWE. In the Thetis framework, \cite{Clare2021} applied MA-based mesh movement in NSWE based hydro-morphodynamic problems such as “Migrating Trench” and wet-dry interface test cases. Although optimal transport based MA mesh movement has significant performance advantages in preventing mesh tangling issues, such robust mesh movement methods lead to a significant increase in computational costs \citep{Clare2021, Song2022}.

With the recent boom in the development of deep learning techniques applied to PDE solvers, there has been increasing interest in using neural networks as learned surrogates for mesh adaptation, aiming to accelerate or replace costly conventional mesh adaptation methods. According to \cite{Song2022}, a series of these methods focusing on mesh generation and mesh refinement have been developed (follow the work of \cite{FIDKOWSKI2021109957}, \cite{huang2021}, \cite{yang2023} and \cite{Zhang2020}), which provide a number of new opportunities and possibilities to address the limitations of traditional mesh adaptation approaches, including high computational costs. To alleviate the mesh tangling and enhance the flexibility of mesh adaptation for traditional MA-based mesh movement, \cite{Song2022} introduced a novel learning-based approach called Mesh Movement Network (M2N) that consists of a Neural Spline model or a Graph Attention Network (GAT). It presented a more direct way to generate the adapted meshes via a neural network compared with the previous AI-based mesh adaptation. As an enhancement and further study of M2N, \cite{zhang2024} introduced Universal Mesh Movement Network (UM2N), which consists of an encoder based on a Graph Transformer and a decoder based on a GAT. Compared to the previous M2N, it selects the element volume loss as training loss instead of the coordinates loss, which is driven by the need for greater physical validity and robustness in automated mesh adaptation. This significantly improves the speed of mesh generation, robustness, and capability of adaptive meshing to address complex boundary problems (following the results in \cite{zhang2024}). However, the complexity of the existing experiments with these methods is still far removed from real-life applications including coastal tsunami modelling.

This work extends research on UM2N \citep{zhang2024} and traditional MA based mesh movement \citep{Andrew2018, Clare2021}
, and further develops non-hydrostatic modelling within the Thetis framework \citep{PAN201968, PAN2021101732}. It also represents the first demonstration of a DG-FE based PDE solver integrated with UM2N (AI for mesh adaptation), applied to non-hydrostatic shallow water equations for the simulation of coastal tsunami propagation and run-up with significant water dispersion. 

The remainder of this paper is organized as follows. Sect.\,\ref{Mathematical} describes the mathematical derivation of the governing equations including the underlying non-hydrostatic ocean model, and the treatment of the wet-dry interface. Advanced discretisation approaches within Thetis' PDE solver and mesh adaptation methods are described in Sect.\,\ref{FE} and Sect.\,\ref{MeshAdapt}, respectively. Sect.\,\ref{Testcase} describes a series of test cases that are used for model verification and validation. Conclusions are presented and results are discussed in Sect.\,\ref{conclusion}.

\section{Mathematical formulation}
\label{Mathematical}
\subsection{Governing equations of non-hydrostatic model}
\label{SWE}
In this study, a single-layer non-hydrostatic free surface model is adopted for our simulations. The total water depth, denoted by $H$\,[\unit{m}], satisfies:
\begin{equation}
\label{eq1}
H = \eta + b,
\end{equation}

\noindent where $\eta$\,[\unit{m}] is the free surface elevation and $b=b(\mathbf{x})$\,[\unit{m}] is the bathymetry.

Although the flow is assumed to be hydrostatic in many ocean models, the effects of wave dispersion can play a significant role in the process of nearshore wave evolution, such as wave propagation on a sloping beach. It is therefore sometimes necessary to consider non-hydrostatic effects caused by water dispersion in the modelling of shoreward wave propagation \citep{PAN201968}. This means that the vertical acceleration may not be ignored. Under the non-hydrostatic condition, the total pressure can be  decomposed into a hydrostatic pressure part $p_{h}$ and a non-hydrostatic part $q$ (cf., e.g., \cite{PAN201968} and \cite{Stelling2003}):
\begin{equation}
\label{eq2}
p = p_{\text{h}} + q,
\end{equation}

\noindent with the hydrostatic pressure written as:
\begin{equation}
\label{eq3}
\frac{\partial p_{\text{h}}}{\partial z} = \rho g
\quad 
\Longrightarrow
\quad
p_{\text{h}} = p_a + \rho g(\eta - z),
\end{equation}

\noindent where $p_a$ is the atmospheric pressure, $\rho$ is the density of the fluid of interest (i.e. sea water) and $g$ is the gravitational acceleration.

The 3D Navier-Stokes momentum equations in non-conservative form can be expressed as:
\begin{equation}
\label{eq4}
\nabla_{\!h}\cdot\mathbf{u}+\frac{\partial w}{\partial z} = 0,
\end{equation}

\begin{equation}
\begin{split}
\label{eq5}
\frac{\partial \mathbf{u}}{\partial t} + \mathbf{u} \cdot \nabla_{\!h}\mathbf{u} + w\frac{\partial \mathbf{u}}{\partial z} &+ F_c\mathbf{e}_z \times\mathbf{u}
+ \frac{1}{\rho}\nabla_{\!h} p \\
&+ \frac{C_d\norm{\mathbf{u}}\mathbf{u}}{H} = \nabla_{\!h}\cdot (\nu\nabla_{\!h}\mathbf{u}) + \frac{\partial}{\partial z}\left( \nu\frac{\partial \mathbf{u}}{\partial z} \right) + \mathbf{S}_{\mathrm{H}},
\end{split}
\end{equation}

\begin{equation}
\label{eq6}
\frac{\partial w}{\partial t} + \mathbf{u} \cdot \nabla_{\!h}w + w\frac{\partial w}{\partial z} + \frac{1}{\rho}\frac{\partial p}{\partial z} + g = \nabla_{\!h}\cdot (\nu\nabla_{\!h}w) + \frac{\partial}{\partial z}\left( \nu\frac{\partial w}{\partial z} \right) + \mathrm{S}_{\mathrm{V}},
\end{equation}

\noindent where $\mathbf{u}$ is the velocity in the horizontal direction and $\nabla_{\!h}$ denotes the horizontal gradient operator. In addition, $F_c$\,[\unit{Hz}], $C_d$(dimensionless) and $\nu$\,[\unit{m^2s^{-2}}] are Coriolis parameter, quadratic drag coefficient and kinematic viscosity respectively. They are used to describe three external and body forces: Coriolis effect, bottom friction and viscosity respectively \citep{Thetis, Wallwork2021}. $\mathbf{S}_{\mathrm{H}}$ denotes other source terms in the horizontal directions such as wind stress and rainfall intensity. Finally, $w$ and $\mathrm{S}_{\mathrm{V}}$ denote velocity and source terms in the vertical direction, respectively.

The kinematic boundary condition at the free surface, $z = \eta(x, y, t)$, and an impermeable land condition at the bottom, $z = -b(x, y)$, are expressed as \citep{PAN201968}:
\begin{equation}
\label{eq7}
w|_{z=\eta} = \frac{\partial \eta}{\partial t} + \mathbf{u}|_{z=\eta}\cdot\nabla_{\!h}\eta,
\end{equation}
\begin{equation}
\label{eq8}
w|_{z=-b} = -\mathbf{u}|_{z=-b}\cdot\nabla_{\!h}b.
\end{equation}

Using the two boundary conditions above, the mass continuity equation in the non-conservative form is obtained via integrating Eq.\,\ref{eq4} over the depth:
\begin{equation}
\label{eq9}
\frac{\partial \eta}{\partial t} +\nabla_{\!h}\cdot (H\mathbf{\Bar{u}}) = 0,
\quad
\text{with}
\quad
\mathbf{\Bar{u}} = \frac{1}{H}\int^\eta_{-b}\mathbf{u}\text{d}z,
\end{equation}

\noindent where $\mathbf{\Bar{u}}$ denotes the depth-averaged horizontal velocity. Then, Eq.\,\ref{eq2} and Eq.\,\ref{eq3} are substituted into Eq.\,\ref{eq5}, and Eq.\,\ref{eq5} is integrated over water depth $H$, the depth-averaged horizontal momentum equation, neglecting vertical advection and viscosity terms, can be written as:
\begin{equation}
\label{eq10}
\frac{\partial \mathbf{\Bar{u}}}{\partial t} + \mathbf{\Bar{u}} \cdot \nabla_{\!h}\mathbf{\Bar{u}} + F_c\mathbf{e}_z \times\mathbf{\Bar{u}} + g\nabla_{\!h}\eta + \frac{1}{\rho H}\int^\eta_{-b}\nabla_{\!h}q\text{d}z = \mathbf{\Bar{F}}_{\!\text{H}} + \mathbf{\Bar{V}}_{\!\text{H}} + \mathbf{\Bar{S}}_{\text{H}},
\end{equation}

\noindent with
\begin{equation*}
\mathbf{\Bar{F}}_{\!\text{H}} = -\frac{C_d\norm{\mathbf{\Bar{u}}}\mathbf{\Bar{u}}}{H} 
\quad
\text{and}
\quad
\mathbf{\Bar{V}}_{\!\text{H}} = \frac{1}{H}\nabla_{\!h}\cdot (H\nu\nabla_{\!h}\mathbf{\Bar{u}}),
\end{equation*}

\noindent where $\mathbf{\Bar{F}}_{\!\text{H}}$ is the bottom friction term, where the coefficient $C_d$ is used to describe a dimensionless bottom friction coefficient. $\mathbf{\Bar{V}}_{\!\text{H}}$ is the horizontal viscosity term, and $\mathbf{\Bar{S}}_{\text{H}}$ represents any other source terms, such as wind stress. Following \cite{PAN201968} and \cite{Stelling2003}, the vertical integral of the non-hydrostatic pressure gradient can be expressed by applying the Leibniz integral rule and linear average approximation:
\begin{align*}
\int^\eta_{-b}\nabla_{\!h}q \, \text{d}z &= \nabla_{\!h}\int^\eta_{-b}q\text{d}z - q|_{z=-b}\nabla_{\!h}b \\
                              &\approx \nabla_{\!h}\left( \frac{1}{2}qH \right) - q|_{z=-b}\nabla_{\!h}b \\
                              &= \frac{1}{2}H\nabla_{\!h}q + \frac{q}{2}\nabla_{\!h}(\eta-b).
\end{align*}

Substituting the approximation above into Eq.\,\ref{eq10}. The depth-average momentum equation in horizontal dimension is:
\begin{equation}
\label{eq11}
\frac{\partial \mathbf{\Bar{u}}}{\partial t} + \mathbf{\Bar{u}} \cdot \nabla_{\!h}\mathbf{\Bar{u}} + F_c\mathbf{e}_z \times\mathbf{\Bar{u}} + g\nabla_{\!h}\eta + \frac{1}{2}\frac{\nabla_{\!h}q}{\rho} + \frac{1}{2}\frac{q}{\rho H}\nabla_{\!h}(\eta-b) = \mathbf{\Bar{F}}_{\!\text{H}} + \mathbf{\Bar{V}}_{\!\text{H}} + \mathbf{\Bar{S}}_{\text{H}}.
\end{equation}

For the vertical dimension, the terms of vertical advection and viscosity are neglected since they are generally small than the vertical non-hydrostatic pressure gradient \citep{PAN201968}. Following \cite{PAN201968} and \cite{Stelling2003}, the vertical velocity $w$ is approximated as a linear profile based on the Keller-box scheme. Therefore, the vertical momentum Eq.\,\ref{eq6} for shallow water can be written as:
\begin{equation}
\label{eq12}
\frac{\partial w}{\partial t} + \frac{1}{\rho}\frac{\partial q}{\partial z} = \frac{\partial}{\partial t}\left( \frac{w|_{z=\eta} + w|_{z=-b}}{2} \right) + \frac{q}{H} = 0.
\end{equation}

\noindent Additionally, one more equation is required to close the system since there are two new variables, $q$ and $w$ added to the system. Hence the 3D continuity Eq.\,\ref{eq4} is expressed as:
\begin{equation}
\label{eq13}
\nabla_{\!h}\cdot\mathbf{u} + \frac{w|_{z=\eta} - w|_{z=-b}}{H} = 0.
\end{equation}

In summary, Eq.\,\ref{eq9} and Eqs.\,\ref{eq11}--\ref{eq13} constitute the governing equations for the depth-integrated non-hydrostatic model. It is noted that Eq.\,\ref{eq11} would yield the hydrostatic NSWE if $q=0$. This means this depth-integrated non-hydrostatic model can be considered the extension of the hydrostatic model under 2D conditions with one layer \citep{PAN201968}. The functionality and transform between non-hydrostatic and hydrostatic are implemented in Thetis \citep{Thetis} and available through a configuration option.

\subsection{Governing equations for a wet-dry domain}
\label{wd}
In many coastal ocean problems, a wetting-drying interface is involved in the simulation of wave propagation, e.g., where a wave generated by tsunami moves to nearshore from a wet domain to a dry domain. Following the wetting-and-drying formulation of \cite{KARNA2011509} and \cite{Thetis} in Thetis, the modified total water depth can be expressed as:
\begin{equation}
\label{eq14}
\Tilde{H} := H + f(H),
\end{equation}

\noindent where $f(H) := (\sqrt{H^2 + \alpha^2} - H) / 2$ is defined as a smooth function to ensure positive water depth. Analogously, the function of bathymetry is also redefined and written as:
\begin{equation}
\label{eq15}
\Tilde{b} = b + f(H).
\end{equation}

Hence, the governing equations of depth-integrated non-hydrostatic model are modified in the following way by taking the wetting-drying effect into account:
\begin{equation}
\label{eq16}
\frac{\partial \eta}{\partial t} + \frac{\partial \Tilde{b}}{\partial t} +\nabla_{\!h}\cdot (\Tilde{H}\mathbf{\Bar{u}}) = 0,
\quad
\text{with}
\quad
\mathbf{\Bar{u}} = \frac{1}{\Tilde{H}}\int^\eta_{-\Tilde{b}}\mathbf{u}\text{d}z.
\end{equation}

\begin{equation}
\label{eq17}
\frac{\partial \mathbf{\Bar{u}}}{\partial t} + \mathbf{\Bar{u}} \cdot \nabla_{\!h}\mathbf{\Bar{u}} + F_c\mathbf{e}_z \times\mathbf{\Bar{u}} + g\nabla_{\!h}\eta + \frac{1}{2}\frac{\nabla_{\!h}q}{\rho} + \frac{1}{2}\frac{q}{\rho \Tilde{H}}\nabla_{\!h}(\eta-\Tilde{b}) = \mathbf{\Bar{F}}_{\!\text{H}} + \mathbf{\Bar{V}}_{\!\text{H}} + \mathbf{\Bar{S}}_{\text{H}},
\end{equation}
\noindent with
\begin{equation*}
\mathbf{\Bar{F}}_{\!\text{H}} = -\frac{C_d\norm{\mathbf{\Bar{u}}}\mathbf{\Bar{u}}}{\Tilde{H}} 
\quad
\text{and}
\quad
\mathbf{\Bar{V}}_{\!\text{H}} = \frac{1}{\Tilde{H}}\nabla_{\!h}\cdot (\Tilde{H}\nu\nabla_{\!h}\mathbf{\Bar{u}}).
\end{equation*}

\noindent The depth-average momentum equation in vertical dimension can be written as:
\begin{equation}
\label{eq18}
\frac{\partial w}{\partial t} + \frac{1}{\rho}\frac{\partial q}{\partial z} = \frac{\partial}{\partial t}\left( \frac{w|_{z=\eta} + w|_{z=-\Tilde{b}}}{2} \right) + \frac{q}{\Tilde{H}} = 0,
\end{equation}

\noindent and Eq.\,\ref{eq13} (Eq.\,\ref{eq6}) is modified as:
\begin{equation}
\label{eq19}
\nabla_{\!h}\cdot\mathbf{u} + \frac{w|_{z=\eta} - w|_{z=-\Tilde{b}}}{\Tilde{H}} = 0.
\end{equation}

In summary, Eqs.\,\ref{eq16}--\ref{eq19} form the governing equations for the depth-integrated non-hydrostatic model with wetting-drying impact. More details of this model can be found in the works of \cite{PAN201968} and \cite{Clare2021} . The main difference between the standard formulation and the wetting-drying type is a mass correction term because of the moving bathymetry (See the second term of Eq.\,\ref{eq16}). This functionality is available through a configuration option in Thetis \citep{Thetis}.

\section{Discretization methods and implementation in Thetis and Firedrake}
\label{FE}
The non-hydrostatic model is solved using the unstructured-mesh finite element coastal ocean modelling system, Thetis \citep{Thetis}, which is built upon the automated code generation framework, Firedrake \citep{FiredrakeUserManual}. Following the pioneering work and relevant extensions of layer-averaged non-hydrostatic approach \citep{Stelling2003, WeiJia} and non-hydrostatic pressure correction \citep{Marshall1997, Stansby1998, Lai2010}, \cite{PAN201968} integrated the layer-averaged non-hydrostatic approach into DG finite element implementation in Thetis. In this formulation the 3D incompressible Navier–Stokes (NS) equations are vertically integrated to obtain a depth-averaged continuity equation for total water depth $H$, and a horizontal momentum equation for the depth-averaged velocity $\mathbf{\Bar{u}}$. Mentioned in Sect.\,\ref{Mathematical}, non-hydrostatic effects are represented by decomposing the pressure into hydrostatic and non-hydrostatic parts and retaining a depth-averaged non-hydrostatic pressure $q$ in the horizontal momentum balance. The vertical momentum and continuity equations are combined to yield a Poisson problem in the depth-integrated \citep{PAN201968}.

In the spatial discretisation, the depth-averaged variables (free-surface elevation and horizontal velocity) are discretised on unstructured triangular meshes with a discontinuous Galerkin (DG) finite element formulation, typically piecewise linear $\textrm{I\!P}^{DG}_1-\textrm{I\!P}^{DG}_1$ spaces. Numerical fluxes across element interfaces are treated using linearised approximate Riemann solvers, and viscosity is stabilised by an interior-penalty formulation \citep{Thetis}. The discrete weak forms are implemented in the high-level Unified Form Language (UFL) within the Firedrake code-generation framework, and the resulting discretised equations are solved using linear solvers available through the PETSc library \citep{PETSc2025, Dalcin2011}. The implementation ensures the non-hydrostatic extension is fully consistent with the existing hydrostatic solver in Thetis, and preserves the properties of high-order convergence and numerical stability of the original hydrostatic DG code (follow the work of \cite{PAN201968}).

Temporal discretisation follows a split-step pressure-correction strategy: in the initial hydrostatic step the free-surface elevation and depth-averaged velocity are advanced using an implicit $\theta$-scheme (i.e., Crank–Nicolson with $\theta = 0.5$) for the free surface and pressure-gradient terms, with advection and other source terms treated explicitly \citep{PAN201968}. In the subsequent non-hydrostatic correction step, the intermediate velocities obtained from the hydrostatic step (not yet divergence-free) are adjusted, which is achieved by solving the Poisson equation for the non-hydrostatic pressure $q$ \citep{PAN201968, CUI20141}. The solution to this equation is then used to correct the horizontal and vertical velocities, effectively projecting them onto a divergence-free field that strictly enforces the 3D continuity equation. These two steps allow the model to accurately represent the characteristics of dispersive waves, while maintaining the stability and efficiency of the underlying hydrostatic framework \citep{PAN201968}.

\section{Universal Mesh Movement Network (UM2N)}
\label{MeshAdapt}
\subsection{Monge–Amp\`{e}re mesh movement}
\label{MAMeshAdapt}
Following the work of \cite{Andrew2018}, \cite{Wallwork2021} and \cite{zhang2024}, the mesh movement process can be defined as the optimization process to find the transformation between a computational domain, $\Omega_C$, and a physical domain, $\Omega_P$. The aim of mesh movement is to obtain the mapping $\mathbf{x}(\boldsymbol{\xi}):\Omega_C \rightarrow \Omega_P$, such that the provided monitor function $m(\mathbf{x})$ satisfies an equidistribution condition. It leads to the optimization problem that can be expressed as \citep{Wallwork2021}:
\begin{equation}
\label{eq46}
\min_{\mathbf{x}:\Omega_C \rightarrow \Omega_P}\Vert \mathbf{x}(\boldsymbol{\xi}, t) - \boldsymbol{\xi}\Vert_{L^2(\Omega_C)}
\quad
\text{subject to}
\quad
m (\mathbf{x}) \det \mathbf{J} = \theta,
\end{equation}

\noindent where $\boldsymbol{\xi}$ are the computational coordinates. $\mathbf{J}$ denotes the Jacobian of the mapping $\mathbf{x}(\boldsymbol{\xi})$, and $\theta$ is a normalisation constant:
\begin{equation}
\label{eq47}
\mathbf{J}_{ij} := \frac{\partial x_i}{\partial\xi_j}
\quad
\text{and}
\quad
\theta := \frac{\int_{\Omega_P}m\text{d}x}{\int_{\Omega_C}\text{d}\xi}.
\end{equation}

According to \cite{Brenier1991} and \cite{DELZANNO20089841}, the optimisation problem Eq.\,\ref{eq46} can be constrained to have a unique solution by using optimal transport theory, where the deformation of the mapping is expressed in the following form:
\begin{equation}
\label{eq48}
\mathbf{x}(\boldsymbol{\xi}) = \boldsymbol{\xi} + \nabla_{\boldsymbol{\xi}}\Phi(\boldsymbol{\xi}),
\end{equation}

\noindent where $\Phi$ denotes a scalar potential. This deformation form, Eq.\,\ref{eq48}, ensures that the quantity $\Phi + \frac{1}{2}|\boldsymbol{\xi}^2|$ is convex,  which in turn guarantees that the mapping is injective \citep{Brenier1991, Andrew2018}. Substituting Eq.\,\ref{eq48} into Eq.\,\ref{eq46}, a nonlinear PDE of Monge–Amp\`{e}re (MA) type is obtained:
\begin{equation}
\label{eq49}
m(\mathbf{x})\det(\mathbf{I} + \mathbf{H}(\Phi)) = \theta, 
\quad
\text{where}
\quad
\mathbf{H}(\Phi)_{ij} := \frac{\partial^2\Phi}{\partial \xi_i \partial \xi_j}
\end{equation}

\noindent is the Hessian of $\Phi$ and $\mathbf{I}$ is the identity matrix.

Notice that the MA type equation (Eq.\,\ref{eq49}) is an auxiliary PDE which is only associated with the process of mesh movement, independent of the governing PDE equations or physical problems to solve \citep{zhang2024}. This means that, although the specific choice of monitor function may be guided by the physical problem (see the following section), the form of the MA equation is invariant across different PDEs. As argued by \cite{zhang2024}, this decoupling enhances the generalization of machine learning based mesh movement methods, i.e., a trained MA neural solver can transfer to new scenarios with a low cost of fine-tuning, or without retraining.

\subsection{Design of Monitor function}
\label{monitor_value}
\subsubsection{Monitor function choice}
Following \cite{Wallwork2021}, the Hessian of a solution field is adopted to constitute the monitor function. This is because it introduces information about curvature, which is crucial for accurately resolving transitions in the solution field. Specifically, curvature information is obtained from the point-wise Frobenius norm of the Hessian $\Vert \mathbf{H}\Vert$ \citep{Clare2021, Wallwork2021}. In this study, the Hessian $\mathbf{H}$ which is used in UM2N is directly computed from the numerical solution of the elevation $\eta_h$. The monitor function based on the Frobenius norm of the Hessian matrix can be written as:
\begin{equation}
\label{eq_monitor}
m := 1 + \alpha\frac{\Vert \mathbf{H}(\eta_h)\Vert}{\max \Vert \mathbf{H}(\eta_h)\Vert},
\quad
\alpha > 0,
\quad
\text{where}
\quad
\Vert \mathbf{H}\Vert = \sqrt{\sum_{i=1}^2\sum_{j=1}^2\mathbf{H}_{ij}}.
\end{equation}

\noindent Here, $\alpha$ is a user-specified parameter, which in this work is set as $\alpha=5$.

Further modifications to the monitor function are introduced in Sect.\,\ref{conical_island_tc}, following the work in \cite{Clare2021}, to better resolve the wetting-drying interface.

\subsubsection{Monitor smoothness}
\cite{Wallwork2021} also mentioned that applying monitor functions based on the Hessian (or gradient) may lead to non-smoothness issues, particularly if the function became too focused on sharp interfaces. To generate adaptive meshes with greater regularity and uniformity, smoothing of the monitor function is performed before mesh movement at each timestep. 

A widely used numerical procedure for the smoothing of a scalar function is through the discretisation of a diffusion equation, e.g. the heat equation:

\begin{equation}
\label{eq51}
\frac{\partial m(\mathbf{x})}{\partial t}=\kappa\nabla^2 m(\mathbf{x}).
\end{equation}

We consider a standard Continuous Galerkin finite element discretisation in combination with an implicit Euler time discretisation. A single (pseudo) time step takes the following form:
\begin{equation}
\label{eq_sm_wk}
\int_\Omega \phi_m \cdot \widetilde{m} \mathbf{d}\mathbf{x}- \int_\Omega \phi_m \cdot m \mathbf{d}\mathbf{x} + \widetilde{\alpha} \int_\Omega (\hat{\mathbf{J}} \cdot \nabla \phi_m) \cdot (\hat{\mathbf{J}} \cdot \nabla \widetilde{m})\mathbf{d}\mathbf{x} = 0.
\end{equation}

\noindent For all test functions $\phi_m$ in the discrete function space $V$, the same as used for discretizing the monitor function $m(x)$, here the space of piecewise linear functions with values at the vertex nodes. We have inserted the finite element Jacobian $\hat{\mathbf{J}}$ of the transformation between reference element and physical element. Here $\hat{\mathbf{J}}$ arises from the finite element formulation and is used to transform the physical space derivative, $\nabla$, to a derivative with respect to reference coordinates. This ensures that the diffusion coefficient $\widetilde\alpha$, which controls the length scale over which the monitor is diffused, is expressed relative to the local grid size. For a constant value of $\widetilde\alpha$ the smoothing occurs over the same number of grid lengths, even where the mesh is anisotropic.

Starting from the given monitor function, e.g., based on \eqref{eq_monitor}, as the initial condition $m(x)$, we perform a single (pseudo) time step by solving \eqref{eq_sm_wk} for $\widetilde{m}$. The value of diffusion coefficient $\tilde\alpha$ is set from 0.3 to 0.5 when performing the monitor smoothing for test cases. In this way, for each timestep $t^{(k)}$, the monitor function $m^{(k)}$ is smoothed to obtain $\widetilde{m}^{(k)}$ which is then fed into the UM2N-based mesh movement.

\subsection{UM2N framework}
\begin{figure}[t!]
    \begin{center}
        \includegraphics[width=1.\textwidth]{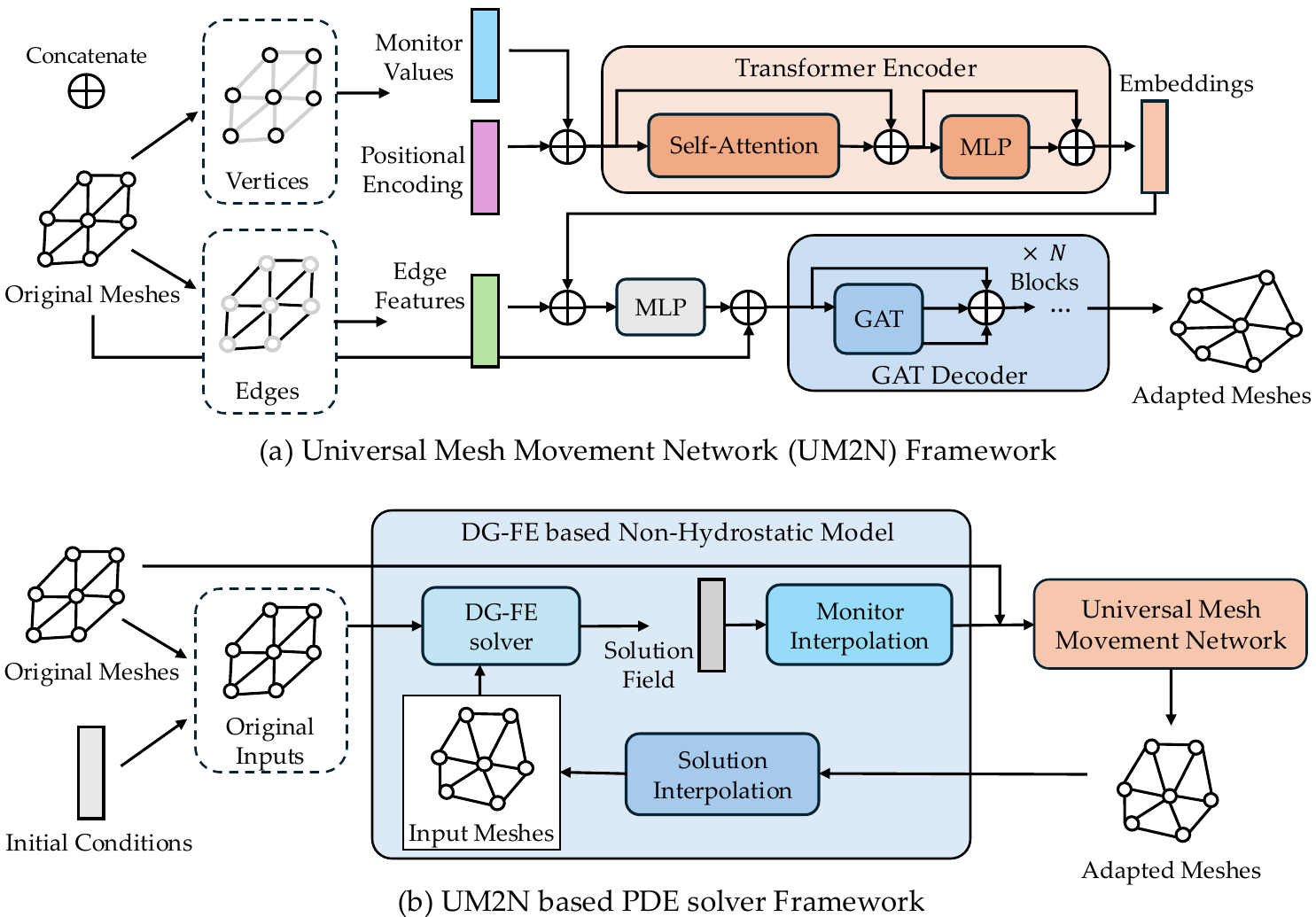}
    \end{center}
    \caption{\label{fig:UM2N_PDEsolver} Overview of Universal Mesh Movement Network (UM2N) shown in (a), cited from \cite{zhang2024}, and UM2N based PDE solver pipeline shown in (b). The monitor interpolation in (b) includes a step of monitor smoothing.}
\end{figure}

The UM2N framework, introduced in \cite{zhang2024}, mainly consists of a Graph Transformer encoder and a Graph Attention Network (GAT) based decoder. The graph transformer encoder is used as a feature extractor. The original mesh coordinates, $\boldsymbol{\xi}$, and monitor values evaluated at these coordinates, $m(\boldsymbol{\xi})$, are set as the input features, in which the coordinates serve as the positional encoding, and features are concatenated and encoded into $\textbf{Q}$\,(Query), $\textbf{K}$\,(Key) and $\textbf{V}$\,(Value) matrices via Multi-Layer Perceptrons (MLPs) for a self-attention module. The embedding $\boldsymbol{z}$ are generated by the encoder as output, which are fed into a subsequent GAT based deformer for mesh movement \citep{zhang2024}.

A Graph Attention Network (GAT) is adopted as the decoder in UM2N, chosen for its ability to restrict vertex movement to within a single hop of neighboring vertices, thereby alleviating mesh tangling issues. The decoder comprises $N$ GAT blocks, where the initial block receives the mesh query $\boldsymbol{\xi}$, the embeddings $\boldsymbol{z}$ from the encoder part, and the edge features $\boldsymbol{e}$ as inputs. For each subsequent $k^{th}$ block, the inputs include the coordinates of the initial mesh $\boldsymbol{\xi}$, the coordinates of the moved mesh $\boldsymbol{\xi}^{(k-1)}$ and the extracted features from the previous $(k-1)^{th}$ block \citep{zhang2024}.

In summary, the input mesh vertex and edge features are collected, with vertex coordinates and monitor function values processed by a graph transformer to generate embeddings. These embeddings, combined with edge features, are fed into a Graph Attention Network (GAT) decoder, which, along with a mesh query, produces the adapted meshes. 

In this work, the specific architecture and training configuration of the UM2N model follow the setup described in~\cite{zhang2024}. The model is trained on a PDE-independent training dataset, $D = \{\boldsymbol{d} : \boldsymbol{d} = (\boldsymbol{\xi}, m_{\boldsymbol{\xi}}; \boldsymbol{V}_{\!r}, \boldsymbol{K}_{\!r})\}$, constructed by original mesh coordinates $\boldsymbol{\xi}$, monitor values $m_{\boldsymbol{\xi}}$, the coordinates of pre-calculated (reference) mesh’s nodes $\boldsymbol{V}_{\!r}$ and elements $\boldsymbol{K}_{\!r}$. This dataset contains $600$ randomly generated samples, each consisting of an original mesh with $463$ or $513$ vertices, the associated monitor function values, and the reference adapted mesh obtained via solving the Monge--Amp\`{e}re equation. It is generated by randomly creating generic solution fields, each formed by summing a random number of Gaussian distribution functions centered at random locations with random widths in different directions to incorporate anisotropy (the details of the full generation procedure is presented in Figure\,A1 and Appendix\,B of \cite{zhang2024}). These generated fields are designed to be interpretable as solutions of arbitrary PDEs. Training is performed using the Adam optimizer on an Nvidia RTX 3090 GPU. Full details of the network hyperparameters (including learning rate, batch size and number of epochs) and dataset generation procedure can be found in the released source code in \cite{zhang2024}.

\subsection{Loss functions}
Following the notation of \cite{zhang2024}, the adapted mesh can be denoted as $\mathcal{H}_a(\boldsymbol{\xi}, m_{\boldsymbol{\xi}}; \boldsymbol{\theta}) = \{ \boldsymbol{V}_{\!a}, \boldsymbol{K}_{\!a} \}$, where $\boldsymbol{\theta}$ is the model parameters of UM2N, $\boldsymbol{V}_{\!a} = \{ \boldsymbol{x}_i \}_{i=1}^{N_1}$ denotes the arrays of vertex coordinates in the adapted meshes, and $\boldsymbol{K}_{\!a} = \{ \boldsymbol{k}_i \}_{i=1}^{N_2}$ denotes the set of elements with $x_i$ as the $i^{\mathrm{th}}$ vertex in the adapted meshes. Similarly, $\boldsymbol{y}_i$, $\boldsymbol{q}_i$ are used to represent the coordinates of the $i^{\mathrm{th}}$ vertex and element respectively in sets $\boldsymbol{V}_{\!r}$ and $\boldsymbol{K}_{\!r}$ for the reference meshes.

Given the dataset $D$, defined in the previous section, the final objective is to find model parameters $\boldsymbol{\theta}$, by minimizing the total loss, $\mathcal{L}$, that consists of the element volume loss, $\mathcal{L}_{vol}$, and Chamfer distance loss, $\mathcal{L}_{cd}$ \citep{zhang2024}. This process can be expressed as:
\begin{equation}
\label{}
\arg\min_{\boldsymbol{\theta}}\mathcal{L}(\boldsymbol{\theta}) := \lambda_{vol}\mathcal{L}_{vol}(\boldsymbol{\theta}) + \lambda_{c d}\mathcal{L}_{cd}(\boldsymbol{\theta}),
\quad
\lambda_{vol}, \lambda_{c d} > 0,
\end{equation}
\noindent where $\lambda_{vol}$ and $\lambda_{c d}$ represent hyper-parameters used to balance these two effects.

The element volume loss is obtained by computing the averaged volume difference between each element in the adapted meshes ($\boldsymbol{K}_{\!a}$) and that in reference meshes ($\boldsymbol{K}_{\!r}$) obtained from the MA movement \citep{zhang2024}. The loss is then expressed as:
\begin{equation}
\label{}
\mathcal{L}_{vol}(\boldsymbol{\theta}) = \mathbb{E}_{(\boldsymbol{\xi}, m_{\boldsymbol{\xi}}; \boldsymbol{V}_{\!r}, \boldsymbol{K}_{\!r})\in D}\Bigg{[}\frac{1}{|\boldsymbol{K}_{\!a}|}\sum_{i=1}^{N_2}|\mathrm{Vol}(\boldsymbol{k}_i) - \mathrm{Vol}(\boldsymbol{q}_i)|\Bigg{]},
\end{equation}
\noindent where $\mathrm{Vol}(\cdot)$ is the volume (area) of a given element.

The Chamfer distance loss is obtained by computing the Chamfer distance that finds the nearest distances for each node between adapted meshes and reference meshes and then sums the square of the distances \citep{zhang2024}. It can be expressed as:
\begin{equation}
\begin{split}
\label{eq_lcd}
\mathcal{L}_{cd}(\boldsymbol{\theta}) = \mathbb{E}_{(\boldsymbol{\xi}, m_{\boldsymbol{\xi}}; \boldsymbol{V}_{\!r}, \boldsymbol{K}_{\!r})\in D}
&\Bigg{[}\frac{1}{|\boldsymbol{V}_{\!a}|}\sum_{\boldsymbol{x}_i\in \boldsymbol{V}_{\!a}}\min_{\boldsymbol{y}_j\in \boldsymbol{V}_{\!r}}\norm{\boldsymbol{x}_i - \boldsymbol{y}_j}_2 \\
&+ \frac{1}{|\boldsymbol{V}_{\!r}|}\sum_{\boldsymbol{y}_j\in \boldsymbol{V}_{\!r}}\min_{\boldsymbol{x}_i\in \boldsymbol{V}_{\!a}}\norm{\boldsymbol{x}_i - \boldsymbol{y}_j}_2\Bigg{]}.
\end{split}
\end{equation}
This metric encourages the model to produce meshes with a spatial distribution of nodes closely resembling that of the reference meshes. Unlike applications in three-dimensional reconstruction or computer vision where Chamfer distance might be calculated by sampling thousands of random points from a continuous surface to compare shapes, UM2N adopts the bidirectional Chamfer distance shown in Eq.\,\ref{eq_lcd}, which operates directly on the discrete vertices of the meshes without requiring additional sampling.

\subsection{UM2N based PDE solver}
\begin{figure}[t!]
    \begin{center}
        \includegraphics[width=1.\textwidth]{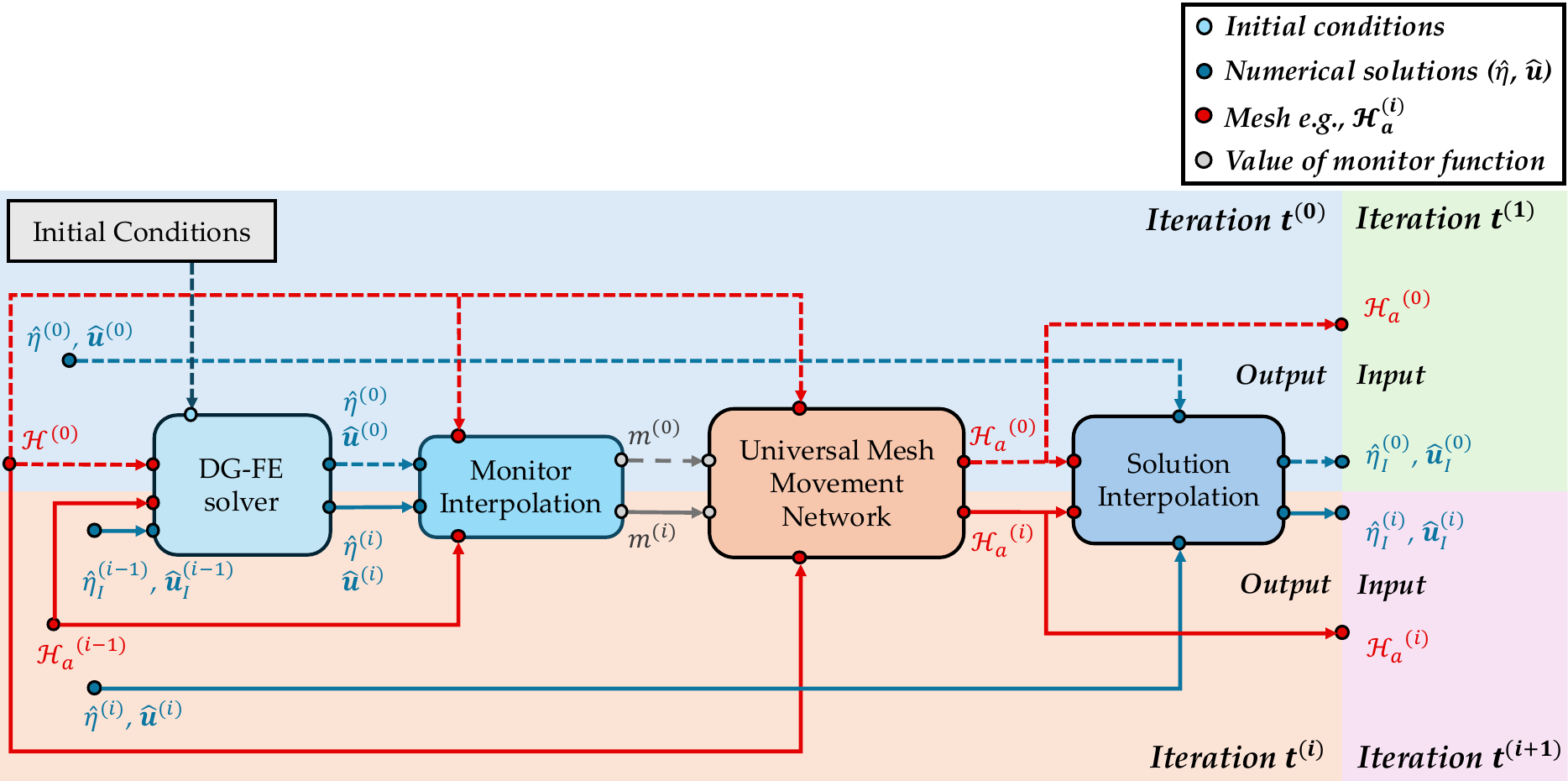}
    \end{center}
    \caption{\label{fig:UM2N_PDE_iteration} The solution procedure of one iteration in the UM2N based PDE solver framework shown in Fig.\,\ref{fig:UM2N_PDEsolver} (b). Blue and green background panels: the solution procedure of one iteration from the initial timestep $t^{(0)}$ to $t^{(1)}$; Orange and purple background panels: the solution procedure of one iteration from the timestep $t^{(i)}$ to $t^{(i+1)}$, where $i=1,2,3, \cdots,n$, $n\in\mathbb{N}$. Top right: Color legends used to represent the inputs and outputs in the solution procedure.}
\end{figure}

Fig.\,\ref{fig:UM2N_PDEsolver} shows the UM2N-based PDE solver framework, which consists of UM2N itself (Fig.\,\ref{fig:UM2N_PDEsolver}\,(a)) which is incorporated in the DG-FE based Non-Hydrostatic solver loop (Fig.\,\ref{fig:UM2N_PDEsolver}\,(b)). The network of UM2N is trained exclusively on the values of the monitor function, which can be computed relatively efficiently from the PDE solution. The desired movement of mesh nodes is determined by solving the Monge–Amp\`{e}re equation for each monitor function, which is derived from randomly generated PDE solutions. Following the training process, the model can be deployed using only the monitor values as input, thereby functioning independently of the specific PDE being solved.

In Fig.\,\ref{fig:UM2N_PDE_iteration}, a schematic diagram further illustrates the solution process of one timestep within the UM2N based PDE solver pipeline. At timestep $t^{(i)}$, one general iteration starts from the adapted mesh $\mathcal{H}_{a}^{(i-1)}$ together with the solution fields solved for in the last time step, including both the free surface elevation and velocity components. These fields are used as input for the DG-FE solver implemented by Thetis \citep{Thetis}, which solves the non-hydrostatic governing equations on $\mathcal{H}_{a}^{(i-1)}$ and produces the updated discrete solutons $\langle\hat{\eta}^{(i)}$, $\hat{\mathbf{u}}^{(i)} \rangle$. A Hessian based monitor function is then computed from the target variable (typically the free surface elevation, see Sect.\,\ref{monitor_value}). The values of the monitor function are further interpolated from the current adapted mesh back onto the original mesh $\mathcal{H}^{(0)}$. The original mesh together with the monitor values $m^{(i)}$ are then fed into UM2N, which predicts the mesh deformation and generates a new adapted mesh $\mathcal{H}_{a}^{(i)}$. Since the newly computed solution fields remain defined on the adapted mesh $\mathcal{H}_{a}^{(i-1)}$, all estimated variables, including the free surface elevation and velocity components, are transferred onto $\mathcal{H}_{a}^{(i)}$ via cross-mesh projection in Thetis/Firedrake, which employs a conservative Galerkin projection based on supermesh construction~\citep{Farrell2009, Farrell2011}, preserving the global integral of the projected fields to machine precision. This yields the transferred solutions $\langle\hat{\eta}_{I}^{(i)}, \hat{\mathbf{u}}_{I}^{(i)}\rangle$. Finally, the updated mesh-solution pair $\langle\mathcal{H}_{a}^{(i)}, \hat{\eta}_{I}^{(i)}, \hat{\mathbf{u}}_{I}^{(i)}\rangle$ is passed to the DG-FE solver for the nest timestep $t^{(i+1)}$, completing one iteration. The initial timestep at $t^{(0)}$ differs only in that the computation starts from the original mesh $\mathcal{H}^{(0)}$ with interpolated initial conditions, rather than from an adapted mesh and the interpolated solutions inherited from a previous timestep. (follow upper blue and green background panels in Fig.\,\ref{fig:UM2N_PDE_iteration}.)

\section{Model Verification and Validation}
\label{Testcase}

\subsection{N-wave strip source}
\label{N-wave}
\begin{figure}[b!]
    \begin{center}
        \includegraphics[width=1.\textwidth]{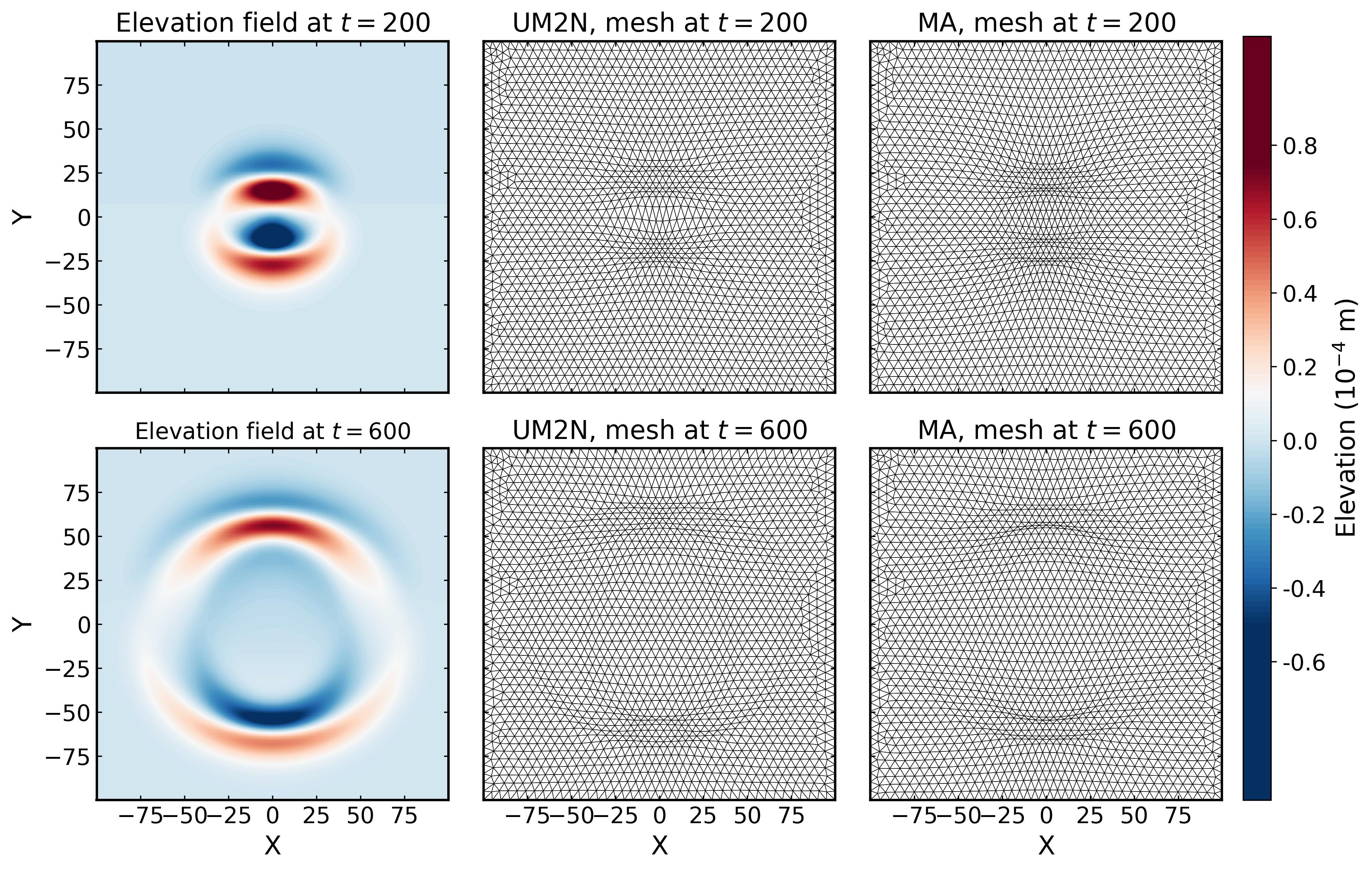}
    \end{center}
    \caption{\label{fig:N_wave_contour} Comparison of N-wave test case: Two-dimensional plan view of wave patterns after propagating at $t=200$ (upper left) and $t=600$ (bottom left), and their corresponding meshes via different mesh configurations: MA movement (middle column) and UM2N (right column). The scale of elevation range for 2D plan view (left column) is from $-0.5\times10^{-4}$ to $0.75\times10^{-4}$ shown by blue to red.}
\end{figure}

\begin{figure}[t!]
    \begin{center}
        \includegraphics[width=1.\textwidth]{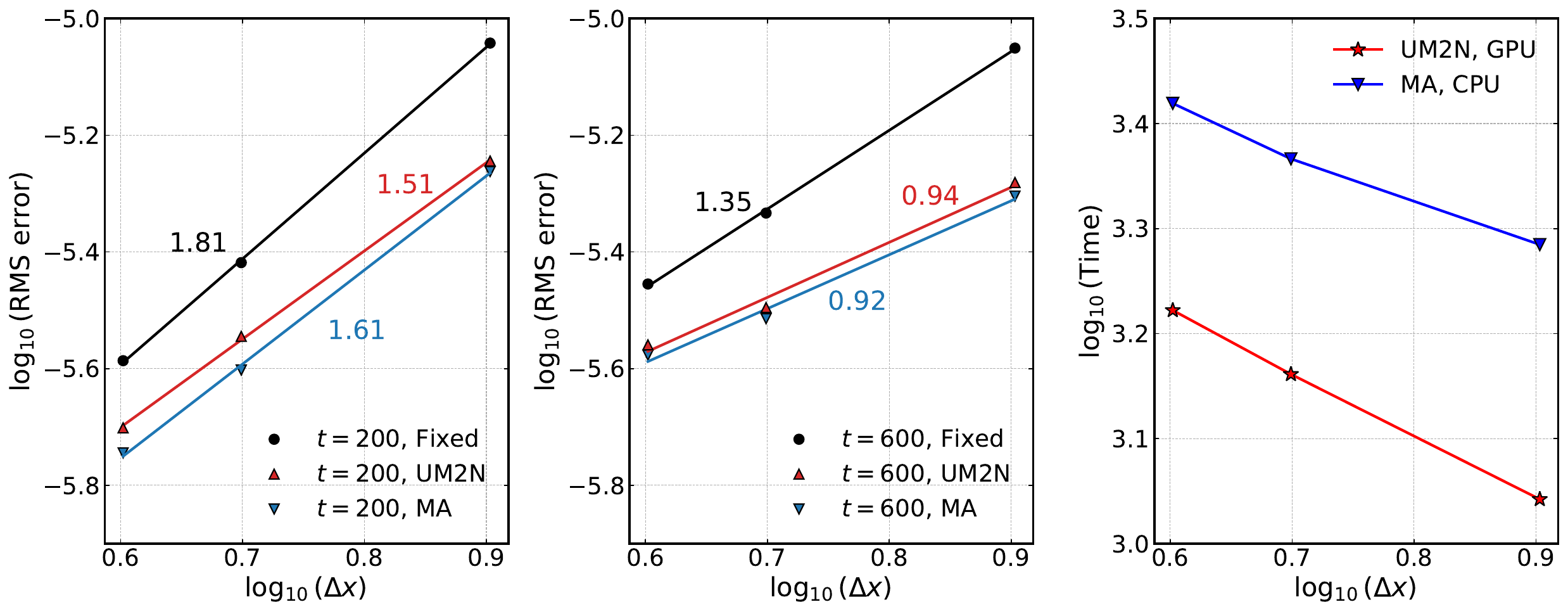}
    \end{center}
    \caption{\label{fig:N_wave_convergence} Convergence of RMS error of free-surface elevations in the N-wave strip source test case for fixed meshes, MA moved meshes and UM2N moved meshes at times $t$=200 (left) and $t$=600 (middle), respectively. Comparisons of computational cost (average runtime of total mesh movement step on CPU/GPU) for MA meshes and UM2N meshes (right). The total mesh movement time includes time spent on monitor calculation and solution interpolation. Mesh sizes correspond to $\Delta x = 8.0$, $5.0$, and $4.0\,\unit{m}$ in convergence plots at left and middle columns, and the plot of computational cost at right column. The numbers show the slope of the least-squares fitted linear lines (black: fixed meshes; blue: MA moved meshes; red: UM2N moved meshes).}
\end{figure}

This problem was introduced in \cite{Kanoglu2013} as an idealised model for tsunami propagation studies, and is based on a finite strip source over a constant bathymetry solved using the linear shallow water equation. This test case is used here to investigate the convergence, robustness and accuracy-cost balance for UM2N, compared to Monge–Amp\`ere (MA) based mesh movement. The computational domain is given by $\Omega = [-100, 100] \times [-100, 100]$ with open boundaries. According to \cite{Kanoglu2013} and \cite{TadepalliSynolakis}, the initial elevation of N-wave type is defined as:
\begin{equation}
\begin{split}
\label{eq59}
\eta_0(x,y) = \frac{\epsilon H}{2}&\left[\tanh{(\gamma_n(x-x_0))} - \tanh{(\gamma_n(x-x_0-L))} \right] \\
&(y-y_2)\text{sech}^2{(\gamma_n(y-y_1))},
\end{split}
\end{equation}

\noindent where $\epsilon$ is a scaling parameter and $H$ is the initial wave height measured from the bottom topography to the free surface. In Eq.\ref{eq59}, $x_0$ and $L$ are the starting point and crest length of the source, respectively. In addition, $y_1$ and $y_2$ are the locations of depression and elevation for an N-wave, respectively. Furthermore, the steepness of the wave is denoted as $p_0$ in the parameter $\gamma_n = \sqrt{3Hp_0/4}$.

In the presented computations, these parameters are set to $\epsilon=-0.04$, $H=0.001$, $x_0=-15$, $L=30$, $y_1=0$, $y_2=2.3$ and $p_0=15$. The time duration of the simulation is $t_{end}=600$. The analytical solution for this test case was derived in \cite{Kanoglu2013}.

Fig.\,\ref{fig:N_wave_contour} presents a comparison of results obtained using the UM2N adaptive mesh strategy against the MA movement for the N-wave strip source test case. The left column illustrates the 2D view of free-surface elevation at simulation times $t$=200 and $t$=600, showing the evolution from an initially compact elevation-depression pair into a distinct, expanding circular wave structure. The corresponding mesh adaptation outcomes for UM2N and MA adaptations at these times are shown in the middle and right columns, respectively. 

At the earlier time $t=200$, the UM2N method primarily mobilizes mesh nodes locally, focusing refinements broadly around areas of corresponding significant wave peaks and troughs, while the MA movement generates tighter refinement regions, distributing nodes in a more structured and uniform manner (see the upper row of Fig.\,\ref{fig:N_wave_contour}). This difference becomes more significant at $t=600$, where UM2N continues to concentrate mesh refinement locally around the immediate vicinity of wave peaks and troughs. Additionally, the deformation of UM2N shows slight refinement in the area between the corresponding wave peaks and troughs, which results in obviously localized and somewhat noisy refinement. Mesh deformation obtained by MA movement continues to concentrate mesh refinement strictly around the immediate vicinity of wave peaks and troughs (follow the bottom row of Fig.\,\ref{fig:N_wave_contour}).

Fig.\,\ref{fig:N_wave_convergence} illustrates the convergence plot of the RMS errors for free-surface elevation in the N-wave strip source test case at two simulation stages, $t$=200 (left) and $t$=600 (middle), comparing fixed meshes, MA moved meshes and UM2N moved meshes with varying mesh resolutions. Despite the qualitative differences identified above between the two adapted meshes, both MA and UM2N mesh adaptation methods demonstrate improved accuracy, achieving lower RMS errors across all mesh resolutions compared to fixed meshes, although the results obtained using the UM2N approach show slightly less accuracy than that of the MA based mesh. Quantitative comparison of computational costs between MA and UM2N is also presented in Fig.\,\ref{fig:N_wave_convergence}. Computational inference time is recorded for a series of numerical experiments with different mesh resolutions. This demonstrates that using the UM2N method provides a $\sim 2\times$ speed-up in the total mesh movement step, which includes time spent on monitor calculation and interpolation steps, compared with conventional MA movement. This means that the UM2N method is able to accelerate the process of mesh movement while achieving accuracy improvement that is comparable with the more costly MA method.

\subsection{Solitary wave over a truncated conical shoal}
\label{shoal}
\begin{figure}[b!]
    \begin{center}
        \includegraphics[width=1.\textwidth]{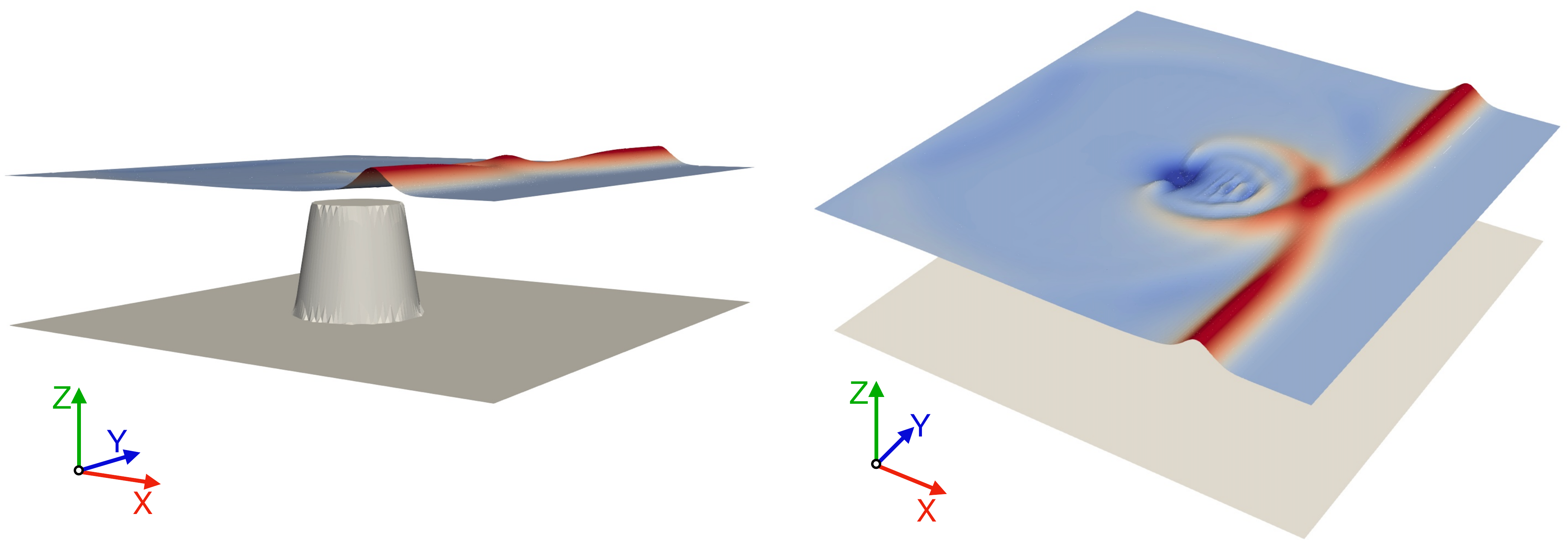}
    \end{center}
    \caption{\label{fig:shoal_dig} Three-dimensional view snapshots of wave patterns after propagating over a truncated conical shoal at $t = 15\,\unit{s}$, from different perspectives.}
\end{figure}

\begin{figure}[]
    \begin{center}
        \includegraphics[width=1.\textwidth]{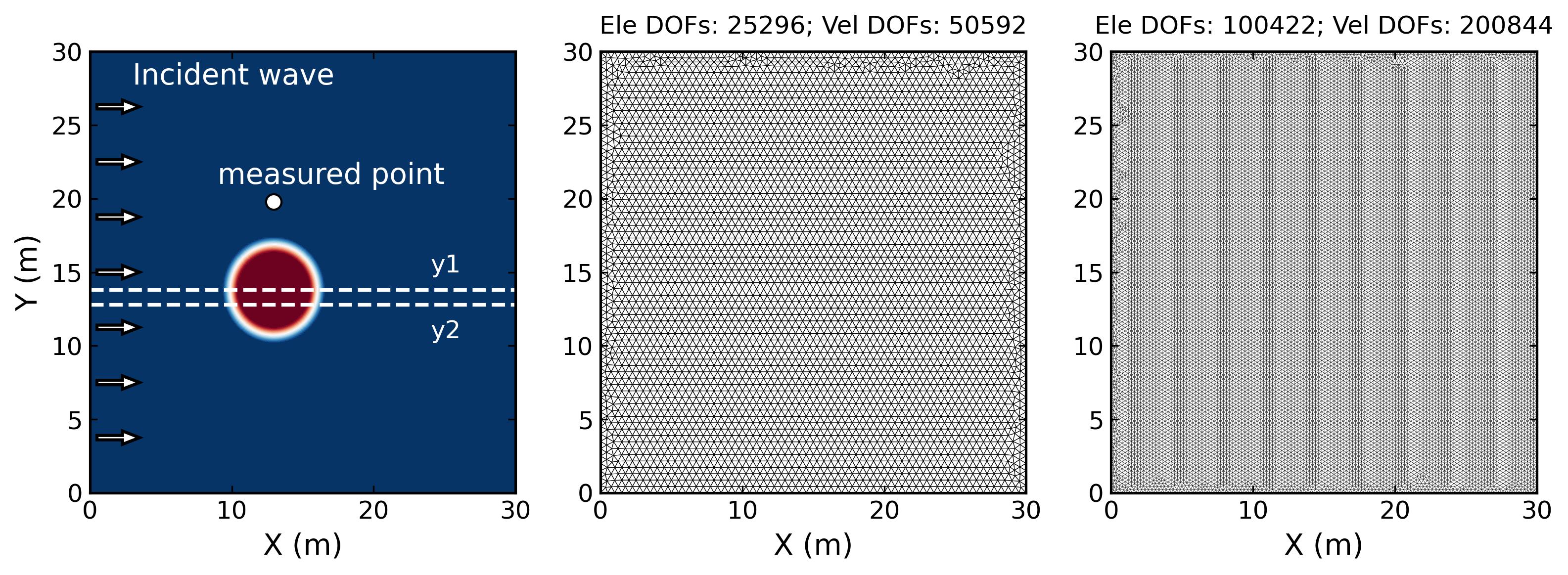}
    \end{center}
    \caption{\label{fig:shoal_seup} Conical shoal setup. Left: Two-dimensional plan view of bathymetry with measurement lines, measurement point (gauge) located at ($x=12.96\,\unit{m}, y=19.80\,\unit{m}$), and diagram
of incident wave ($y_1 = 13.80\,\unit{m}$, $y_2 = 12.80\,\unit{m}$). Middle: coarse mesh topology. Right: fine mesh
topology.}
\end{figure}

\begin{figure}[t!]
    \begin{center}
        \includegraphics[width=1.\textwidth]{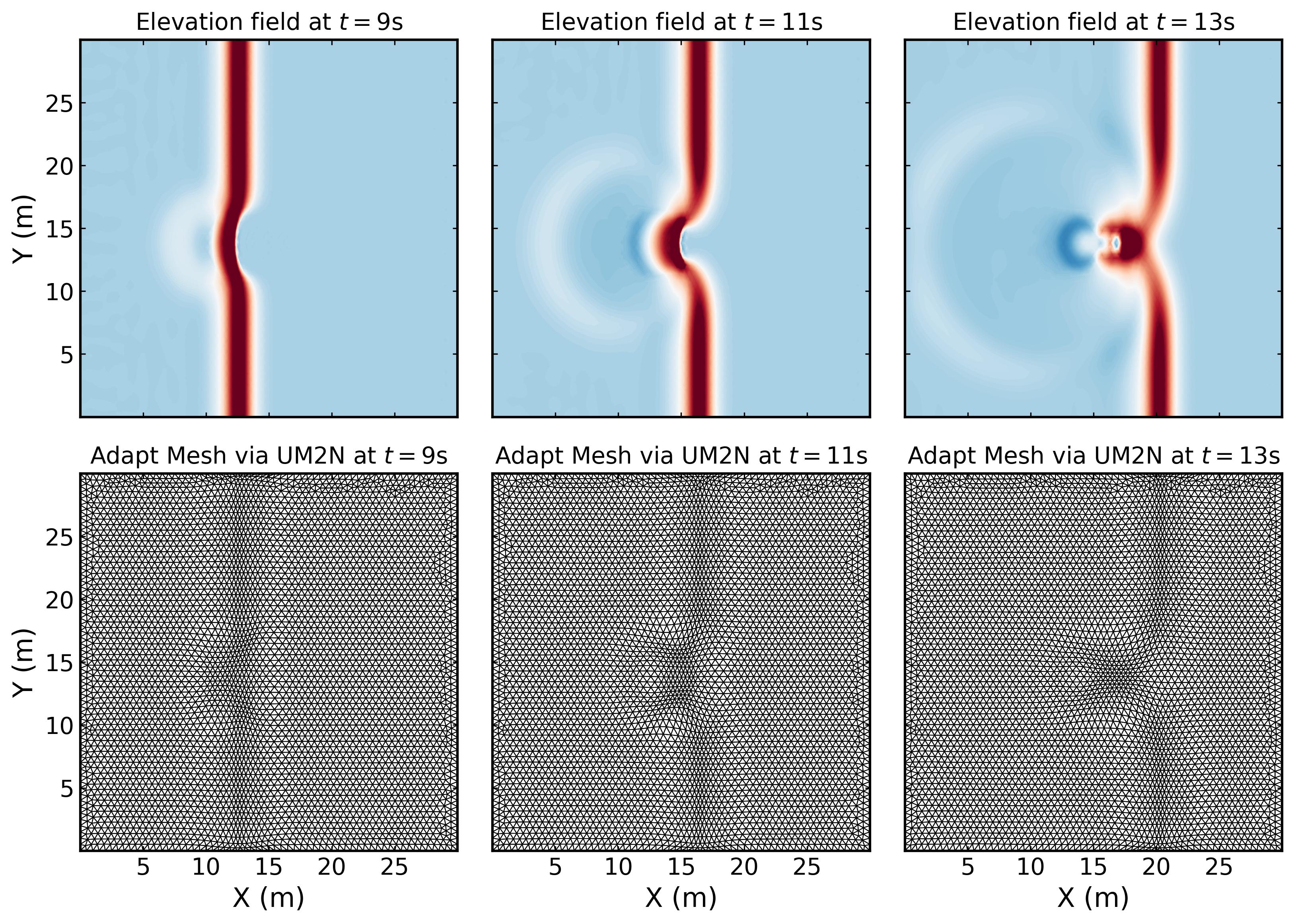}
    \end{center}
    \caption{\label{fig:shoal_mesh} Top: Two-dimensional plan view of wave patterns when solitary wave passes the shoal at different time instants, $t=9.0, 11.0, 13.0$\,\unit{s}. Corresponding adapted mesh using UM2N, performed on coarsed mesh. The scale of free surface elevation range is from $-0.02$ to $0.04$ shown by blue to red.}
\end{figure}

\begin{figure}[t!]
    \begin{center}
        \includegraphics[width=.9\textwidth]{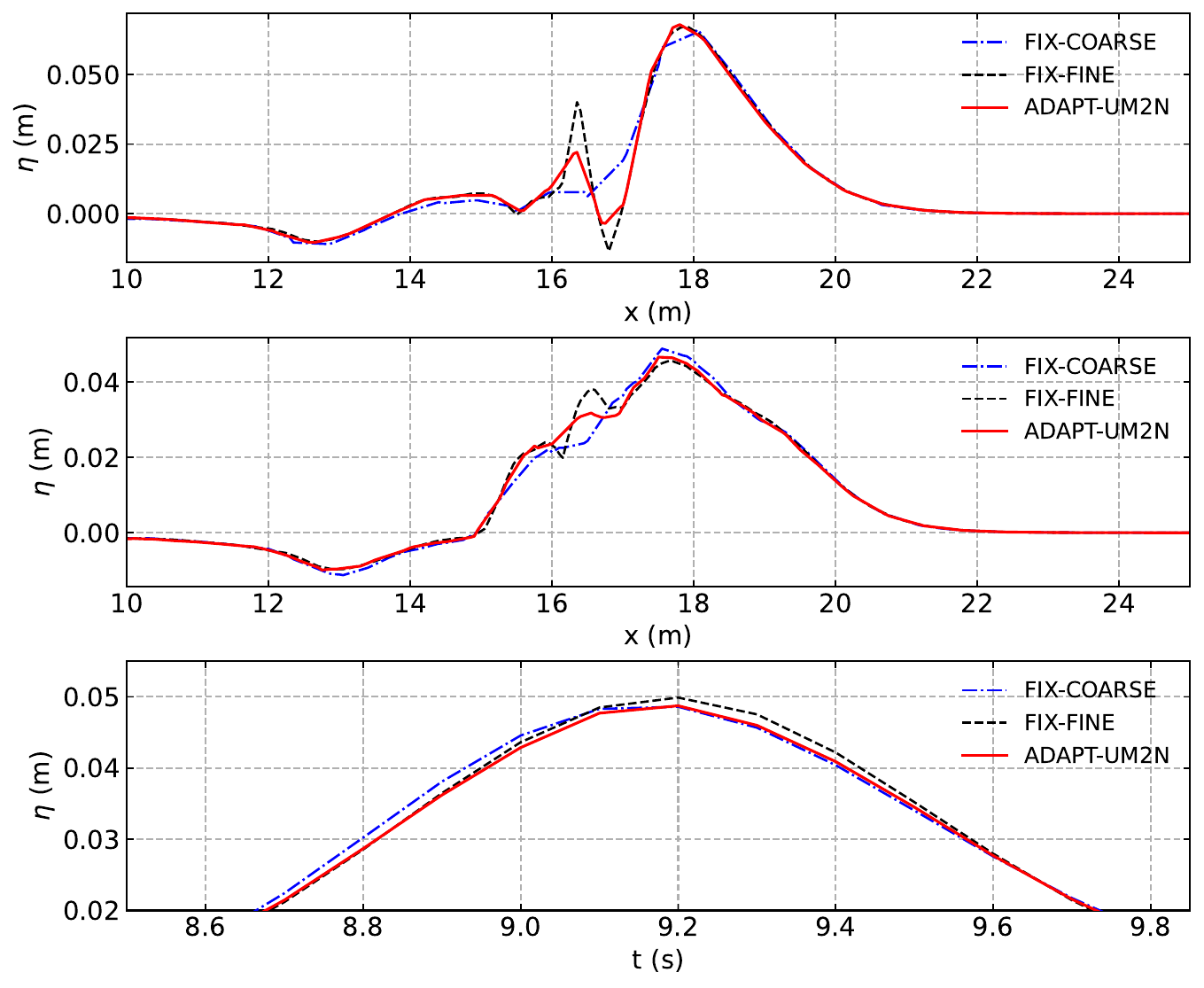}
    \end{center}
    \caption{\label{fig:shoal_elev} Numerical results comparison of results using UM2N on coarse mesh (red solid line), fixed coarse mesh computations (blue dashed line) and fixed fine mesh (black dashed line). Free-surface elevation in the cross sections at $t=13\,\unit{s}$ are presented in the top and middle rows (Top: $y_1$-cross session; Middle: $y_2$-cross session). Bottom: time series of free-surface elevation at fixed measured point.}
\end{figure}

This test case is designed to investigate our framework's abilities in simulating solitary wave run-up and refraction using the UM2N approach over a non-flat, wet bed bottom topography (see Fig.\,\ref{fig:shoal_dig}). The computational domain is set as $[0, 30]\text{\,\unit{m}} \times [0, 30]\text{\,\unit{m}}$ (see Fig.\,\ref{fig:shoal_seup}). A truncated conical-shaped shoal is located at coordinates (12.96\,\unit{m}, 13.80\,\unit{m}). The shoal has a $0.25$\,\unit{m} height, and is submerged by $0.32$\,\unit{m}-depth water initially. The non-flat part of bottom topography is characterized by a truncated conical shoal, featuring a top diameter of 2.60\,\unit{m} and a base diameter of 3.60\,\unit{m}. According to \cite{NIKOLOS20093723} and \cite{HOU2013132}, the left boundary is set as the wave inflow boundary where an input solitary wave with elevation $\eta$ is defined by:
\begin{equation}
\label{eqsw}
\eta (t) = A\text{sech}^2\left[\sqrt{\frac{3A}{4d^3}}C(t-T)\right],
\end{equation}
\noindent where $A$ denotes the amplitude of the wave and $C:=\sqrt{g(A+d)}$ is the wave celerity. A 5\,\unit{m}-wide sponge layer is located at the right outlet boundary. In this numerical experiment, the ratio $A/d = 0.181$ is selected for numerical validation and comparison of the MA and UM2N approaches. The resolution of the coarse mesh is $\Delta x = 0.5\,\unit{m}$ with Elevation degree of freedom count (DOFs) totaling $25,296$ and Velocity DOFs $50,592$, while the resolution of the fine mesh being $\Delta x = 0.25\,\unit{m}$ with Elevation DOFs totaling $100,422$ and Velocity DOFs $200,844$.  The results obtained via both fixed coarse mesh and fixed fine mesh have been used as a reference for those obtained on the coarse mesh (Elevation DOFs $= 100,422$, and Velocity DOFs $= 200,844$), with mesh adaptation approaches (i.e., UM2N and MA). The time step is set to $\Delta t = 0.01\,\unit{s}$.

Fig.\,\ref{fig:shoal_mesh} presents the 2D contour of elevation and corresponding adapted mesh when the solitary wave passes the shoal at different time instants, $t=9.0, 11.0, 13.0$\,\unit{s}. Note that the traditional MA method fails (i.e., the solver diverges) within 70$\sim $100 time steps in this scenario, due to divergence in solving the MA equation. In general, UM2N based movement is able to track the incoming and refracting waves robustly over the whole time period considered. It shows agreement in capturing the whole solitary wave when passing the shoal.

Numerical comparisons between results of UM2N, fixed
coarse mesh computations and fixed fine mesh are shown in Fig.\,\ref{fig:shoal_elev}. We find that the UM2N based adaptive simulation is able to better resolve free-surface behaviour in the cross sections than the fixed coarse mesh, especially in capturing the wave peaks (see top and middle rows in Fig.\,\ref{fig:shoal_elev}). Additionally, the time series elevation at the fixed measurement point indicates that UM2N based movement provides results that are closer to the fine mesh results while the fixed coarse mesh results deviate slightly in the timing of wave peak. This indicates that UM2N adaptive mesh also outperforms the fixed coarse mesh in tracking the independent solitary wave outside the shoal region.

\subsection{Solitary wave over a conical island}
\label{conical_island_tc}
\begin{figure}[b!]
    \begin{center}
        \includegraphics[width=1.0\textwidth]{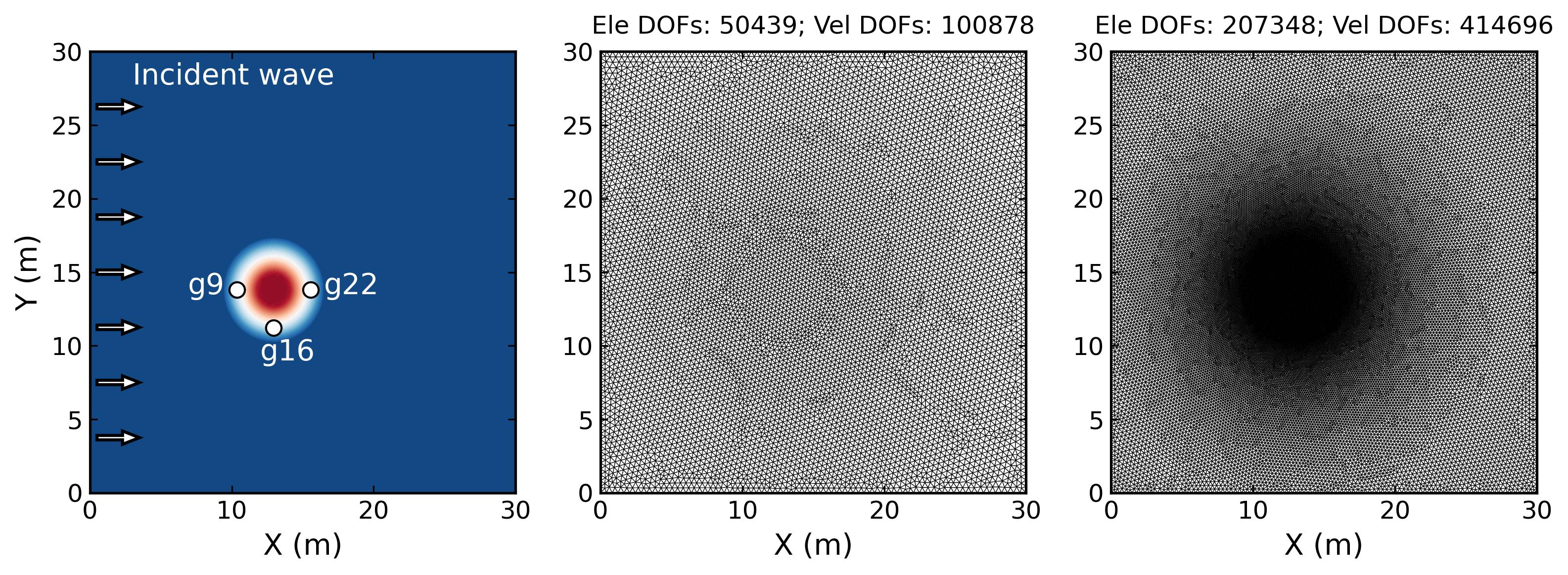}
    \end{center}
    \caption{\label{fig:setup_conical} Conical island setup. Left: Two-dimensional plan view of bathymetry with gauges and diagram of incident wave, where the locations of three gauges, g9: $\text{Gauge 9} = (10.36\,\unit{m}, 13.80\,\unit{m})$, g16: $\text{Gauge 16} = (12.96\,\unit{m}, 11.22\,\unit{m})$ and g22: $\text{Gauge 22} = (15.56\,\unit{m}, 13.80\,\unit{m})$, are presented respectively. Middle: coarse mesh topology for numerical experiment. Right: fine mesh topology for numerical experiment. Ele DOFs: free-surface Elevation Degree of Freedoms; Vel DOFs: Velocity Degree of Freedoms.}
\end{figure}
\begin{figure}[b!]
    \begin{center}
        \includegraphics[width=1.\textwidth]{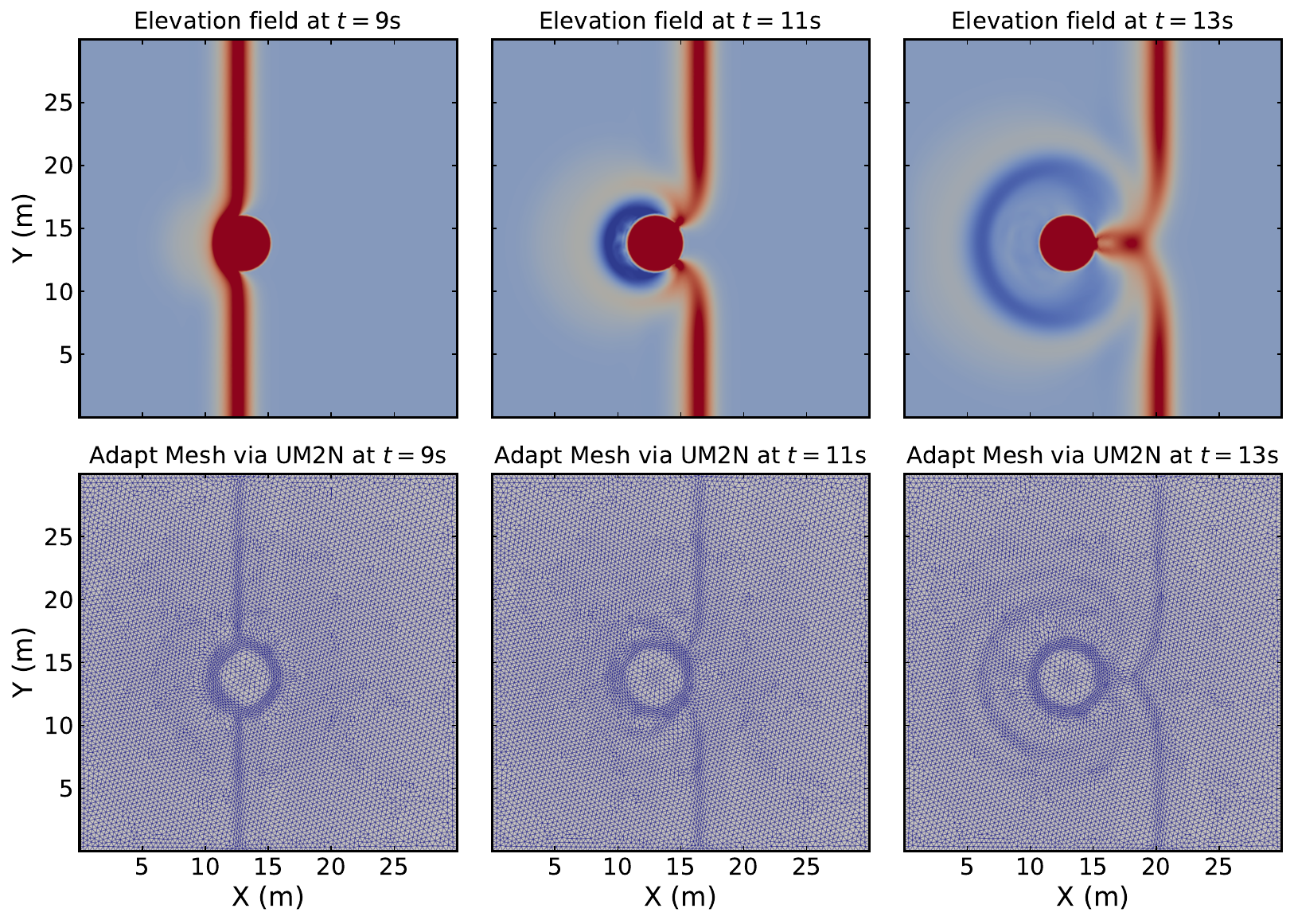}
    \end{center}
    \caption{\label{fig:mesh_conical} Conical island: Two-dimensional plan view snapshots of wave patterns (Figures in the upper row) and corresponding adapted mesh obtained via UM2N (Figures in the bottom row) at different time instants, $t=9.0, 11.0, 13.0$\,\unit{s}. The scale of free surface elevation range in the 2D plan view is from $-0.02$ to $0.04$ shown by blue to red.}
\end{figure}

\begin{figure}[t]
    \begin{center}
        \includegraphics[width=1.\textwidth]{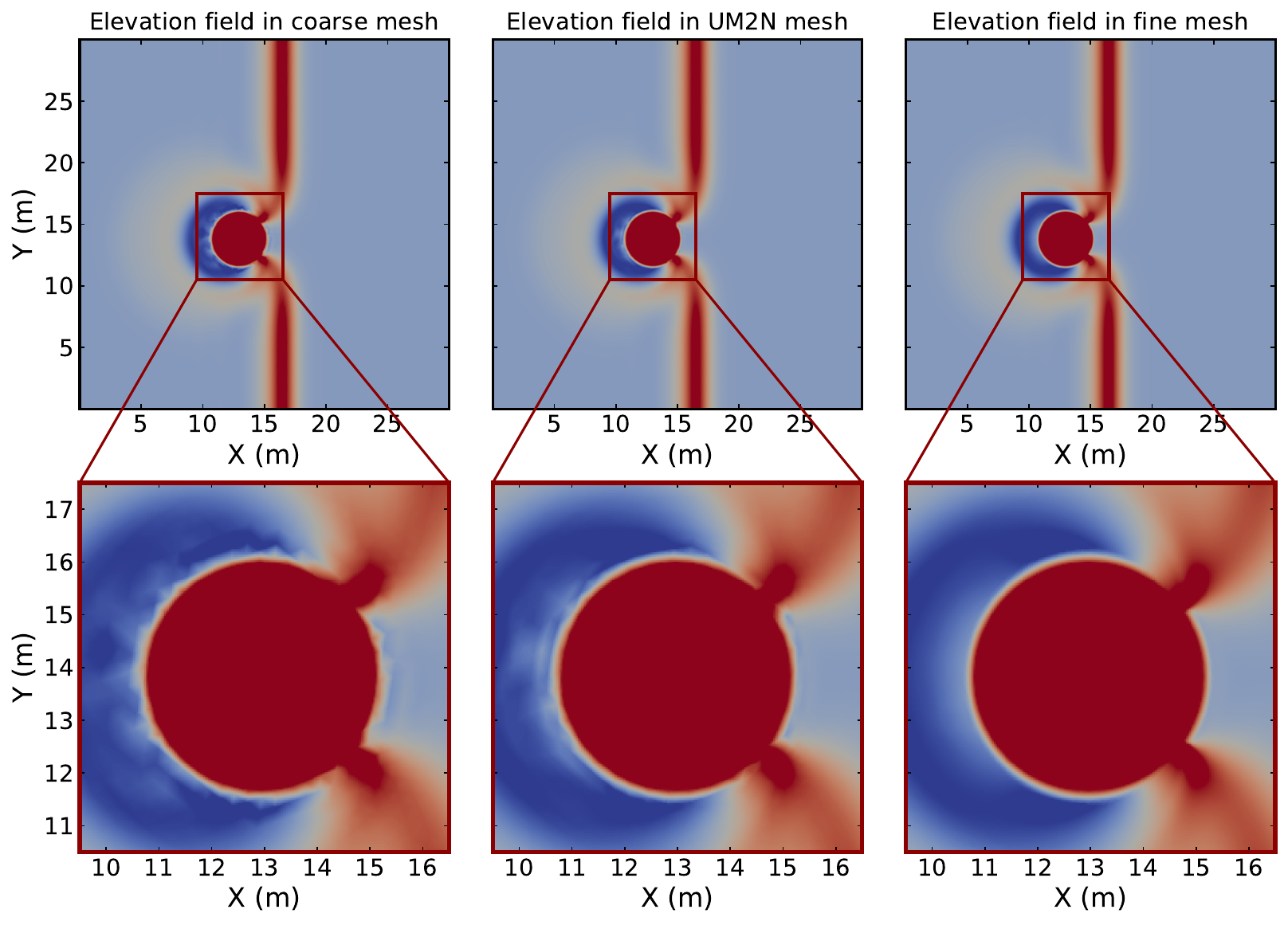}
    \end{center}
    \caption{\label{fig:refined_conical} Conical island: Two-dimensional plan view snapshots of elevation field when the wave propagates over conical island at $t=11.0\,\unit{s}$ obtained via different meshes. Left: Fixed coarse mesh. Middle: UM2N adapted mesh (performed on the same coarse mesh used on the left). Right: Fixed fine mesh. The scale of free surface elevation range is from $-0.02$ to $0.04$ shown by blue to red.}
\end{figure}

\begin{figure}[t!]
    \begin{center}
        \includegraphics[width=1.\textwidth]{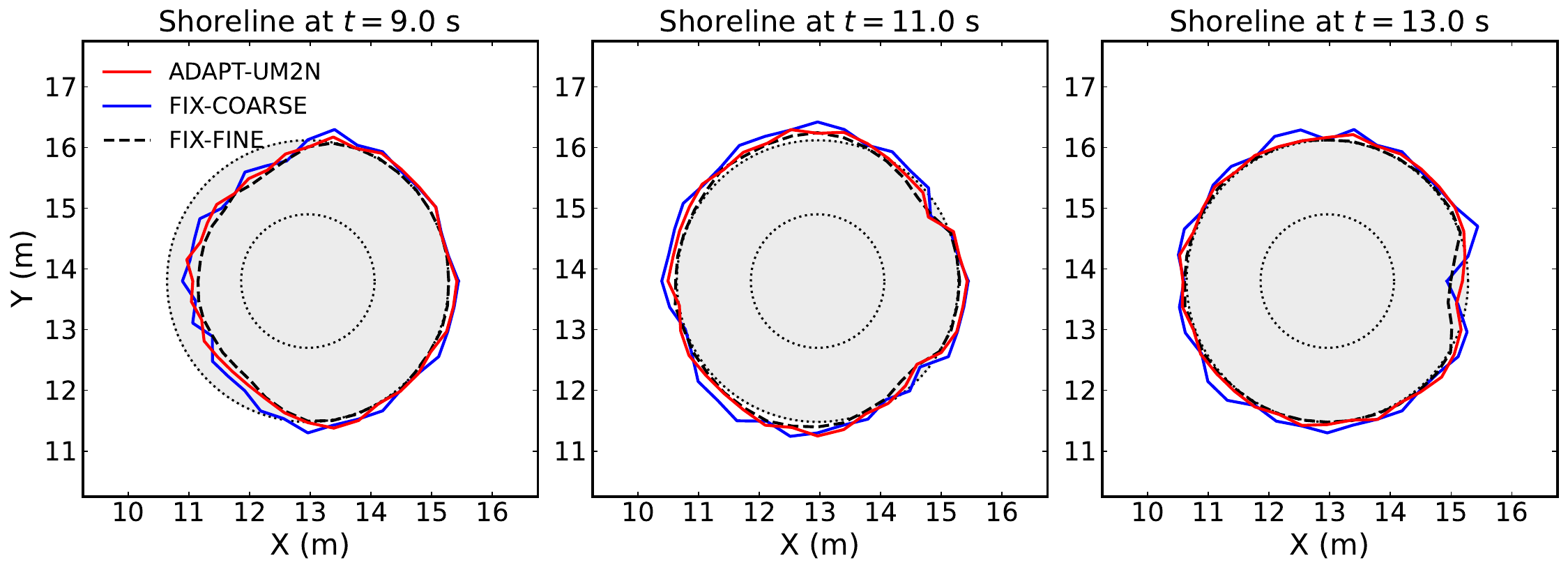}
    \end{center}
    \caption{\label{fig:runup_conical} Shoreline of Conical island obtained on fixed coarse mesh (blue solid line), fixed fine mesh (black dashed line) and UM2N mesh (red solid line) at different time instants, $t=9.0, 11.0, 13.0$\,\unit{s}. The inner and outer black, dotted circles indicate the top of island ($r = 1.1\,\unit{m}$) and initial waterline ($r = 2.32\,\unit{m}$), respectively.}
\end{figure}

\begin{figure}[b!]
    \begin{center}
        \includegraphics[width=.565\textwidth]{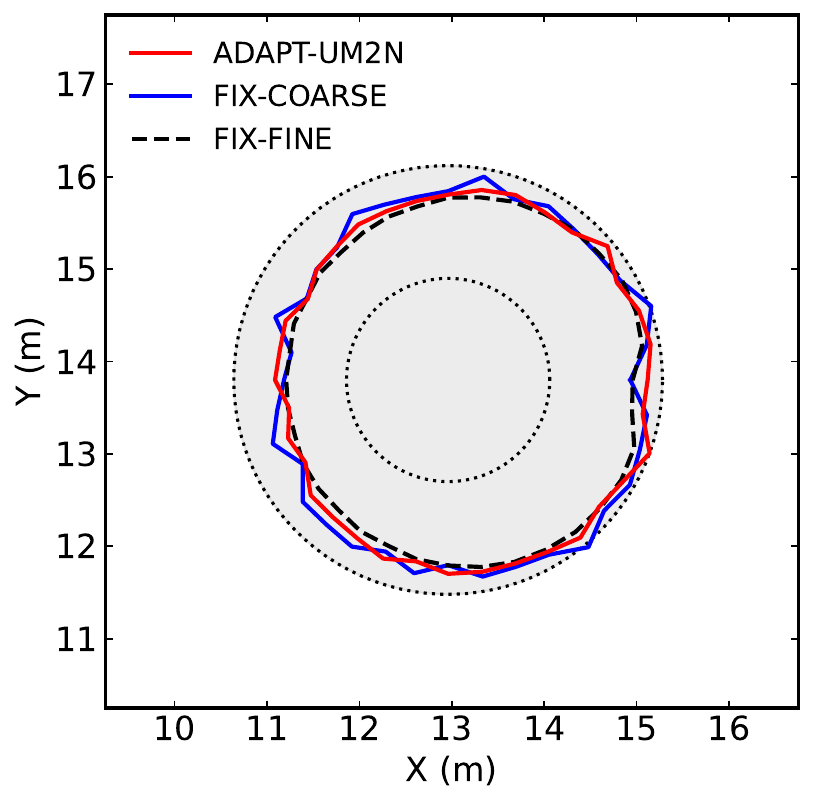}
    \end{center}
    \caption{\label{fig:maxrunup_conical}Conical island: Maximum run-up outlines accumulated over the full simulation period for fixed coarse mesh (blue solid line), fixed fine mesh (black dashed line) and UM2N mesh (red solid line). The inner and outer black, dotted circles indicate the top of island ($r = 1.1\,\unit{m}$) and initial waterline ($r = 2.32\,\unit{m}$), respectively.}
\end{figure}

\begin{figure}[t!]
    \begin{center}
        \includegraphics[width=.9\textwidth]{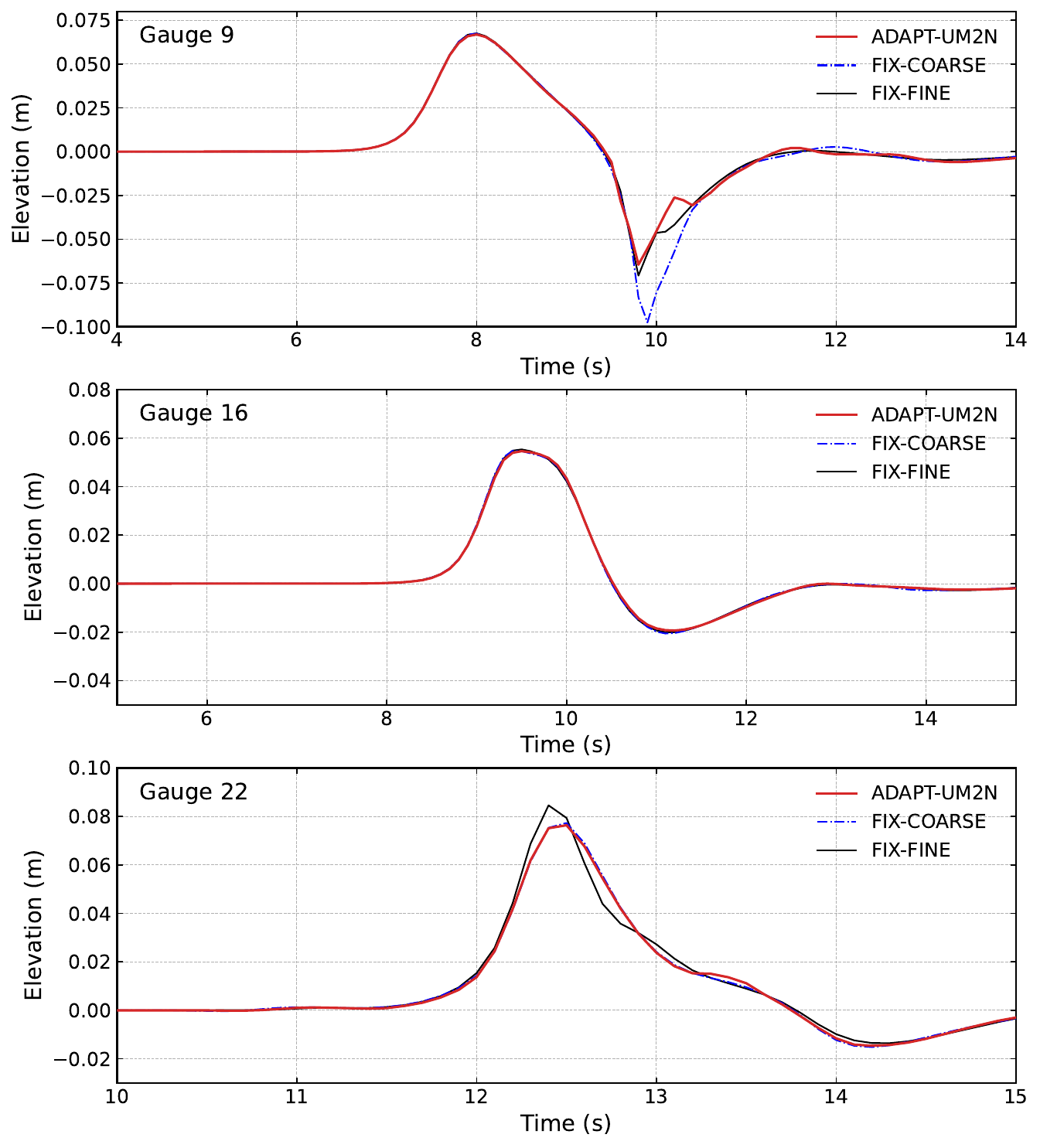}
    \end{center}
    \caption{\label{fig:time_elev_conical} Comparisons of numerical results between adaptive mesh approaches and fixed mesh computations. Free-surface elevation Time series are compared among results of UM2N (red solid line), fixed coarse mesh computations (blue dashed line) and fixed fine mesh (black solid line) at three gauges for the case with $A/d = 0.181$. Gauge locations correspond to those in the original laboratory experiment of~\cite{Briggs1995}: $\text{Gauge 9} = (10.36\,\unit{m}, 13.80\,\unit{m})$, $\text{Gauge 16} = (12.96\,\unit{m}, 11.22\,\unit{m})$ and $\text{Gauge 22} = (15.56\,\unit{m}, 13.80\,\unit{m})$.}
\end{figure}

This classical test case was introduced as a laboratory experiment and is used here to test the performance of each mesh adaptive approaches in terms of tracking wave run-up, refraction and reflection, under wetting-drying interface conditions (see \cite{Briggs1995}, \cite{LYNETT200289}, \cite{HOU2013132}, \cite{ARPAIA2018175} and \cite{PAN201968}). A truncated conical island is situated within a basin measuring 30\,\unit{m} long and 30\,\unit{m} wide. The island has a 0.625\,\unit{m} height, with its center located at coordinates (12.96\,\unit{m}, 13.80\,\unit{m}), and it is submerged by 0.32\,\unit{m}-depth water at the beginning. Its geometry is characterized by a truncated cone, featuring a top diameter of 2.2\,\unit{m} and a base diameter of 7.2\,\unit{m}. There is the same initial elevation with solitary wave type setup and boundary conditions (See Eq.\,\ref{eqsw}) as the previous case.

We select the experiment with $A/d = 0.181$ as benchmark for numerical validation and comparison of the MA and UM2N approaches. A wet-dry interface occurs within the region of the initially submerged island, with significant gradients in both wave elevation and velocity fields resulting around the island. Hence, the configuration of the coarse meshes is such that higher mesh resolution is used with an element size $\Delta x = 0.3$\,\unit{m} in the vicinity of the island, while $\Delta x = 0.5$\,\unit{m} is adopted elsewhere, resulting in the number of elevation degree of freedoms (DOFs) being $50,439$. Similarly, in this experiment, results obtained via both fixed coarse mesh and fixed fine mesh are used as a reference for those obtained via mesh adaptation approaches (i.e., UM2N and MA), on the coarse mesh with the same DOFs/mesh resolutions. The time step is set to $\Delta t = 0.01$\,\unit{s}. The coefficient of bottom friction is assumed to be $0.01$, and horizontal viscosity is set as $0.015$.

Instead of using the monitor function as in previous test cases\,\ref{N-wave}--\ref{shoal}, a component that is designed to track the wetting-drying interface is added. We also set the movement conditions via the monitor function in order to ensure that the mesh refinement has capability in tracking the movement of wave moves across the wet-dry interface and will not undergo additional redundant refinement at the dry domain when the wave passes over the wet-dry interface. The new monitor function is given by

\begin{equation}
\label{}
m_\alpha(x, y) = 
\begin{cases}
1 + \mu \left(\alpha \frac{\min(\|H(\Tilde{\eta})\|, H_{\max})}{H_{\max}} +\frac{\lambda}{\cosh(b_\lambda(\eta-b))^2}\right)
     &\quad\text{if } \eta > -b, \\
1  &\quad\text{otherwise.} \\ 
\end{cases}
\end{equation}

\noindent where $\eta$ is the free surface elevation, $\Tilde{\eta} = \eta + f(H)$ is modified elevation accounting for the treatment of the wet-dry interface mentioned in Section\,\ref{wd}, $b$ is denotes as the bathymetry, $b_\lambda$ controls the width of the wet-dry interface tracker (see \cite{Clare2021}), and the parameters $\mu$, $\alpha$ and $\lambda$ are all user-defined parameters. In the new monitor function, moreover, $H_{\max} = p_{\textrm{cap}}\max\|H(\Tilde{\eta})\|$ is an enforced capping of $\max\|H(\Tilde{\eta})\|$ to enhance the local refinement of the solitary wave, where $p_{\textrm{cap}}\in (0, 1)$ is also a user-defined parameter.

Qualitatively, in general, the proposed UM2N method can be observed to move the mesh in a manner that captures the fluid dynamics of the entire domain, which is shown in Fig.\,\ref{fig:mesh_conical}. Around the conical island, it can be observed that vertices are shifted to improve resolution, resolving the free-surface elevation as expected (See Fig.\,\ref{fig:mesh_conical} and Fig.\,\ref{fig:refined_conical}).

To further quantify the spatial improvement provided by UM2N near the island, the instantaneous shoreline positions and maximum run-up outlines are compared across the three mesh configurations in Figs.\,\ref{fig:runup_conical} and \ref{fig:maxrunup_conical}.  Fig.~\ref{fig:runup_conical} indicates the wetting-drying boundary at $t=9.0, 11.0, 13.0\,\unit{s}$. At $t=9.0\,\unit{s}$ when the solitary wave starts ascending the island, the shoreline positions on coarse mesh and on UM2N mesh are broadly similar. As the wave refracts and runs up the island at $t=11.0\,\unit{s}$, the fixed coarse mesh under-predicts the inundation area on the front face (location at $x\approx10.5\,\unit{m}$) of the island compared with the fine mesh reference, while the result of UM2N tracks the fine-mesh shoreline more closely. At $t=13.0\,\unit{s}$, when the refracted waves converge on the lee side (location at $x\approx15\,\unit{m}$), the UM2N shoreline remains in closer agreement with the fine mesh reference, while the coarse mesh shoreline shows visible deviations. Fig.\,\ref{fig:maxrunup_conical} presents the maximum run-up outlines accumulated over the full simulation. The RMS errors of maximum run-up position are calculated for both fixed coarse mesh and UM2N mesh. Using the solution of the fixed fine mesh as reference, the RMS error of the maximum run-up is $3.185\times 10^{-2}\,\unit{m}$ for the fixed coarse mesh and $2.107\times 10^{-2}\,\unit{m}$ for the UM2N mesh, corresponding to a $33.86\,\%$ reduction in RMS error achieved by UM2N relative to the fixed coarse mesh. These results demonstrate that UM2N method achieves a measurable improvement in resolving run-up and in tracking the wetting-drying interface around the island.

Fig.\,\ref{fig:time_elev_conical} shows comparisons between time series of free-surface elevations from mesh adaptive approaches and the fixed mesh computations. In general, it can be seen that the results of the non-hydrostatic model align closely with the available experimental data, which is consistent with the results shown in \cite{PAN201968}. The results obtained on the three gauges show that UM2N movement provides results comparable in capturing the wave peak to those obtained on both coarse and fine mesh. These indicate that there is little influence of the mesh size in simulating the wave peak at these gauges. Despite of these results, UM2N method improve the free surface elevation in tracking the de-shoaling interference between the two refracted waves in the lee of the island (when the wave propagates around $t=10\,\unit{s}$ at Gauge 9), compared with the results on fixed coarse mesh (see Gauge 9).

Note that we intended to compare UM2N to traditional MA mesh movement for this scenario. However, the MA method fails (i.e., the solver diverges) in about $100$ time steps. The primary reason is that the iterative solver for the nonlinear Monge--Amp\`{e}re equation fails to converge. In this test case, the wetting-drying interface and wave refraction around the island produce very steep gradients in the monitor function. When the monitor function varies by very large factors, the relaxation method used in MA mesh movement may lead to divergence according to \cite{Andrew2018}. Additionally, convergence is further hindered by abrupt changes in the monitor function between time-steps as the solitary wave propagates~\citep{weller2016, Budd2009}. Similarly, failures of a MA solver have been reported by~\cite{zhang2024} in a T\={o}hoku tsunami simulation using Firedrake~\citep{FiredrakeUserManual} framework, where the MA equation failed to converge near complex coastline geometries. Although sufficient numerical results are unavailable for a direct comparison between UM2N and MA method in this scenario, the robustness of UM2N is clearly demonstrated, which entirely bypasses the iterative MA solver.

\subsection{Real-world/laboratory study case: Monai Valley}
\label{monaivalley_testcase}
\begin{figure}[t!]
    \begin{center}
        \includegraphics[width=1.\textwidth]{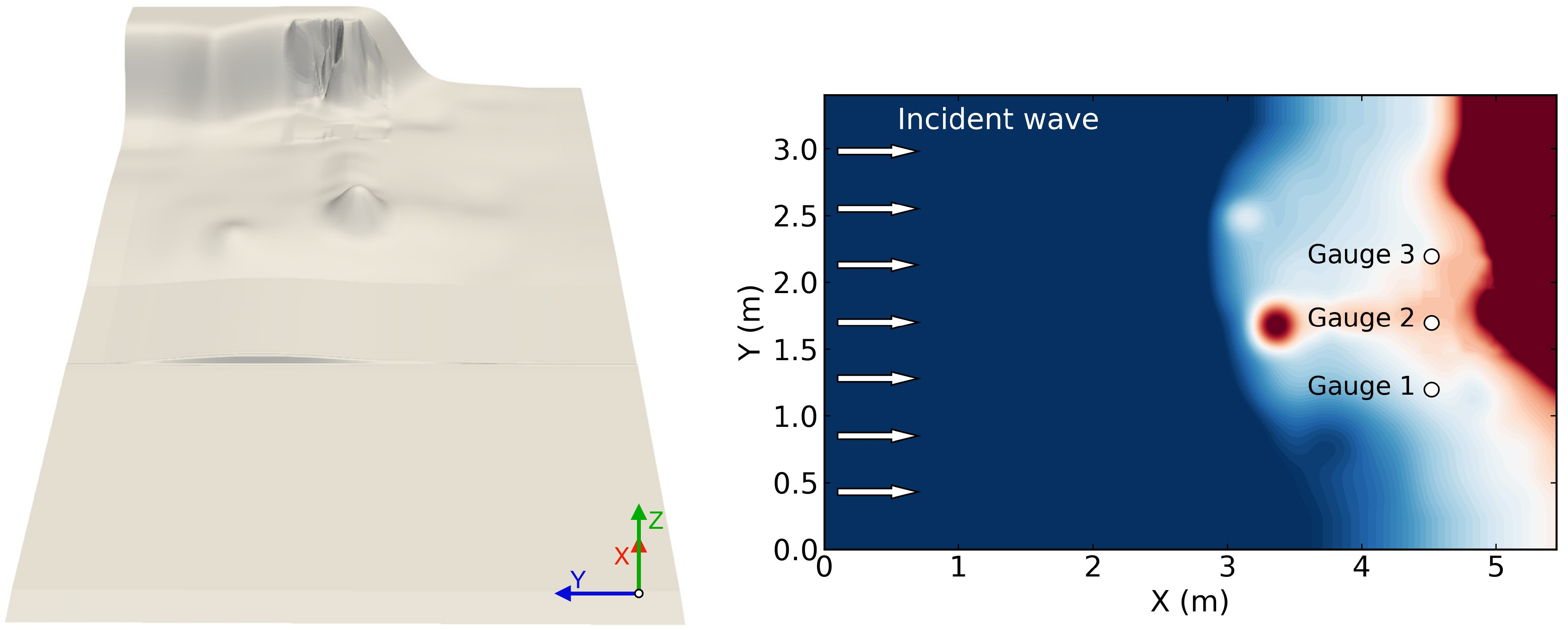}
    \end{center}
    \caption{\label{fig:mv_setup} Monai Valley test case. Left: Bottom topography and land. Right: Setup diagram of the numerical experiment. The location (x, y) of the three gauges are $\text{Gauge 1} = (4.521\,\unit{m}, 1.196\,\unit{m})$, $\text{Gauge 2} = (4.521\,\unit{m}, 1.696\,\unit{m})$, and $\text{Gauge 3} = (4.521\,\unit{m}, 2.196\,\unit{m})$ shown in setup diagram repectively.}
\end{figure}

\begin{figure}[b!]
    \begin{center}
        \includegraphics[width=1.\textwidth]{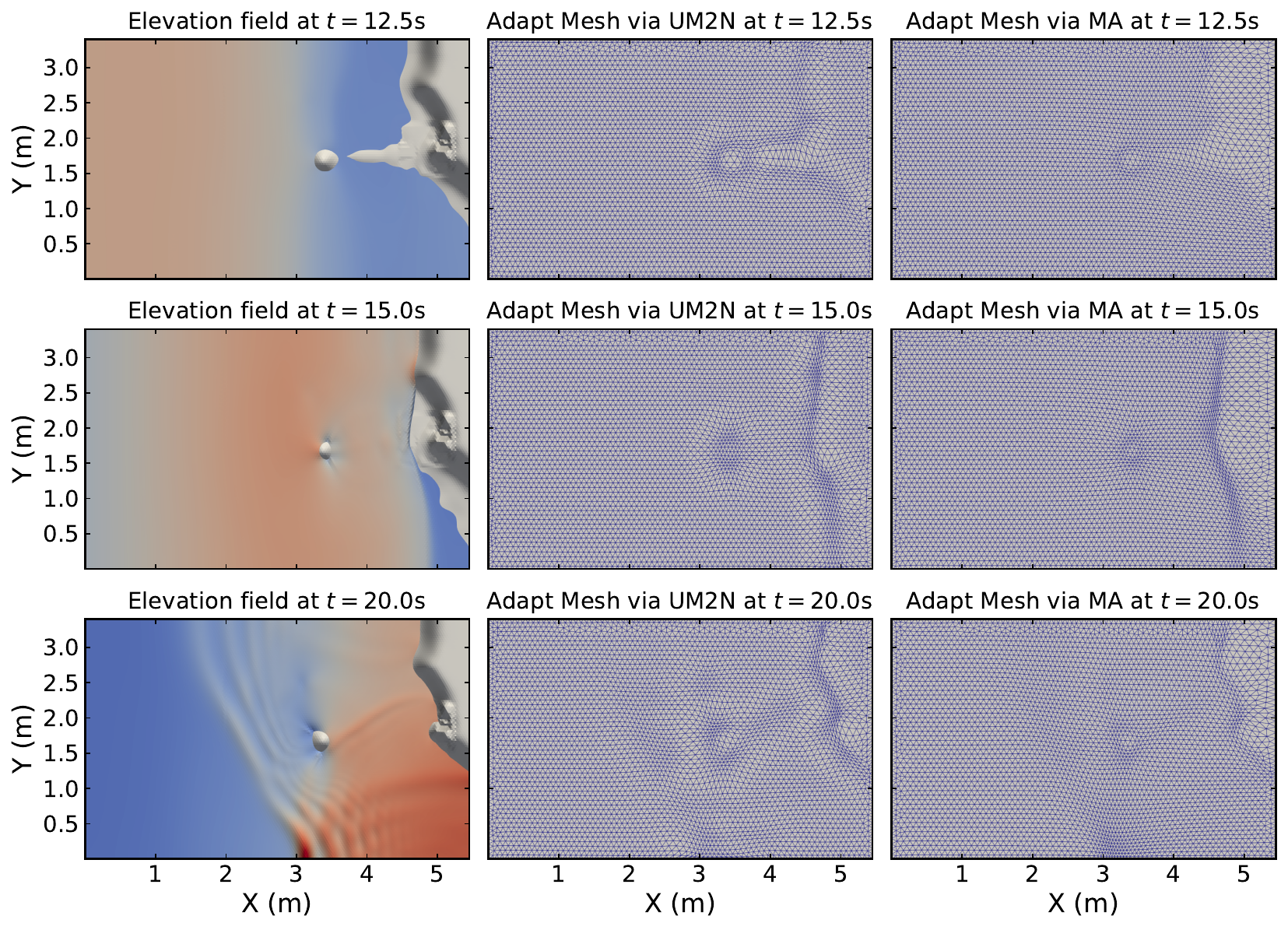}
    \end{center}
    \caption{\label{fig:mesh_mv} Monai Valley: Two-dimensional plan view snapshots of elevation field (left column) under wet-dry interface and adapted mesh using UM2N (middle column) and MA movement (right column) at different time instants, $t=12.5, 15.0, 20.0\,\unit{s}$ (from upper to bottom row, respectively). The scale of free surface elevation range is from $-0.02$ to $0.04$ shown by blue to red.}
\end{figure}

\begin{figure}[t!]
    \begin{center}
        \includegraphics[width=.75\textwidth]{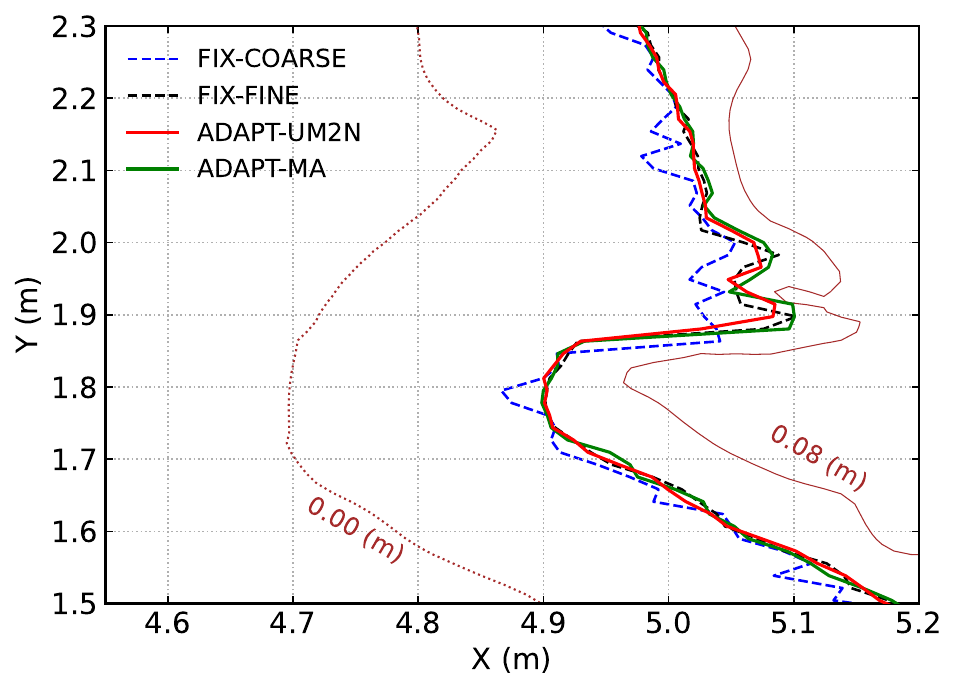}
    \end{center}
    \caption{\label{fig:maxrunup_mv_compare} Maximum run-up outlines accumulated over the full simulation period for different mesh configurations. The outlines (contours) are compared between the results using a fixed coarse mesh (blue dashed line), fixed fine mesh (black dashed line), UM2N mesh (red solid line), and MA mesh (green solid line). The brown outlines indicate the initial shoreline ($0.00\,\unit{m}$, dotted line) and the $0.08\,\unit{m}$ land elevation contour (solid line).}
\end{figure}

\begin{figure}[t!]
    \begin{center}
        \includegraphics[width=1.\textwidth]{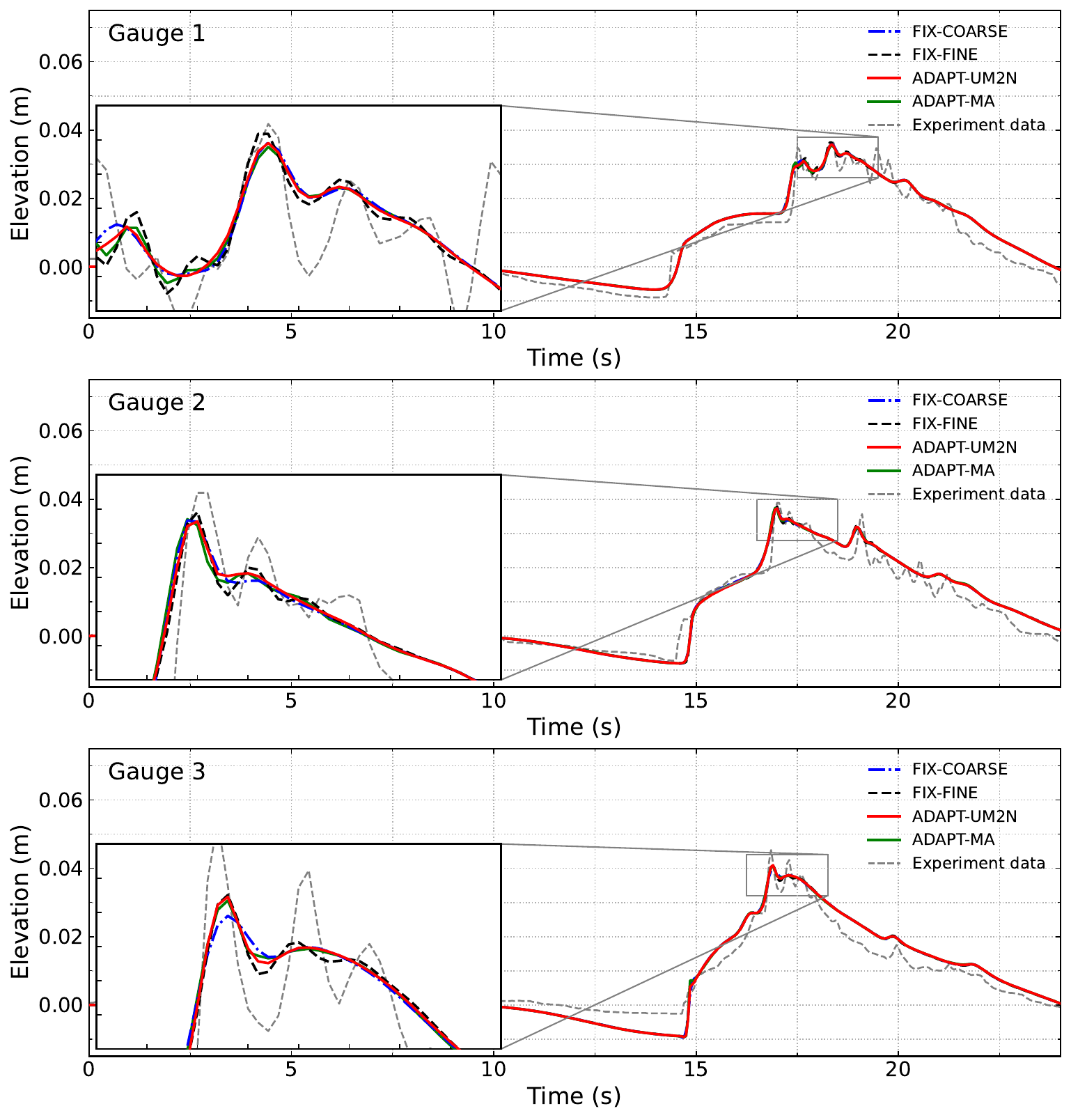}
    \end{center}
    \caption{\label{fig:elev_mv} Comparisons of free-surface elevation time-series at three gauges in the Monai Valley test case. Free-surface elevation time series are compared among results of UM2N (red solid line), MA movement (green solid line), fixed coarse mesh (blue dashed line) and fixed fine mesh computations (black solid line).}
\end{figure}

The following case study focuses on tsunami runup in Monai Valley, Japan. This is a 1/400 scaled laboratory experiment carried at the Central Research Institute for Electric Power Industry (CRIEIPI) in Abiko, Japan, which use a 205\,\unit{m} long, 3.4\,\unit{m} wide and 6\,\unit{m} deep tank in this laboratory experiment. In the numerical experiment, this lab benchmark case is confined to a nearshore region measuring $5.448\,\unit{m} \times 3.402\,\unit{m}$, as a computational domain. Initially, the tank is filled with still water, and a known incident wave moves to the shoreline from the offshore.

In this simulation, the coarse mesh is generated with $4,021$ vertices and $7,802$ elements, with mesh sizes $\Delta x = 0.075\,\unit{m}$; the fine mesh contains $17,953$ vertices and $35,396$ elements, with mesh sizes $\Delta x = 0.035\,\unit{m}$. The bottom friction in this simulation is set as $\mu=0.019\,\unit{s\cdot m^{-1/3}}$ via 2D quadratic drag parameter in the term of bottom friction in the Eq.\,\ref{eq17}, and the magnitude of maximum horizontal velocity is set as $5.0\,\unit{m\cdot s^{-1}}$. Initially, the total time duration is set as $24\,\unit{s}$, to minimize the impact of reflection from the computational boundary, and the time step is $0.01\,\unit{s}$. In addition, the given data of input wave is interpolated into the meshes and is updated at each timestep. The elevation data obtained in the laboratory experiment is compared with numerical results at the three locations: Gauge\,1: $(x_1, y_1) = (4.521\,\unit{m}, 1.196\,\unit{m})$, Gauge\,2: $(x_2, y_2) = (4.521\,\unit{m}, 1.696\,\unit{m})$, and Gauge\,3: $(x_3,y_3) = (4.521\,\unit{m}, 2.196\,\unit{m})$, corresponding to the point $(x_0, y_0) = (0,0)$ shown in Fig.\,\ref{fig:mv_setup}.

The qualitative impact of the two mesh-adaptation strategies, UM2N and MA movement, is presented in Fig.\,\ref{fig:mesh_mv}. In our implementation, UM2N concentrates mesh very locally around steep free-surface gradients, producing highly refined regions near wave peaks and troughs, while relatively irregular mesh sizes and shapes away from these features. Compared with UM2N, the strategy of MA movement redistributes mesh nodes more globally, leading to smoother transitions in mesh size and improved mesh regularity and keeping enhanced resolution in the neighborhood of the wave. These observations indicate that the UM2N strategy is more aggressive in capturing wave features in a fine scale by localized adaptation, while MA movement offers a balanced way between local refinement and overall mesh quality. (see Fig.\,\ref{fig:mesh_mv}).

To further quantify the spatial improvement provided by the mesh adaptation methods near the shoreline, Fig.\,\ref{fig:maxrunup_mv_compare} shows the maximum run-up outlines over the full simulation, compared across different mesh configurations. Along the central part of the shoreline (location at $y\in(1.8\,\unit{m}, 2.1\,\unit{m})$), where the tsunami inundation is most significant, the UM2N outline closely follows both MA and fine mesh results. Moreover, Table\,\ref{tab:maxrunup_mv} presents the quantitative results of run-up error on the given mesh configurations. Using the fixed fine mesh result as reference solution, UM2N achieves a $31.82\,\%$ reduction in RMS error of maximum run-up along the shoreline, which is comparable to the $44.69\,\%$ reduction obtained by the MA method. In addition, Fig.\,\ref{fig:elev_mv} shows the quantitative comparisons in impact of the UM2N method for free-surface elevation time-series at three gauges. Although there is little influence of the mesh resolution on the given gauge locations mentioned by \cite{RICCHIUTO2015306} and \cite{ARPAIA2018175}, the results at Gauge 3 show that the UM2N and MA movement computations approach closer to that of the fine mesh, compared to the results of the coarse mesh. Table\,\ref{tab:maxrunup_mv} indicates that UM2N achieves a larger error reduction ($91.14\,\%$) than the traditional MA movement ($73.64\,\%$) in simulating wave peak at Gauge 3. 

Regarding computational costs, Table\,\ref{mv_benchmark} presents the total runtime, inference time, and mesh movement time for UM2N and MA mesh movement. On GPU, UM2N achieves a $31.83\,\%$ reduction in total runtime and a $\sim 2.09\times$ speedup in total mesh movement time  relative to the MA movement. This includes time spent on interpolation, monitor calculation and smoothing, which for the MA method is deeply integrated in the iterative process and therefore cannot be separated from the core mesh movement time. Using UM2N these costs are incurred as a separate preprocessing step before the actual machine-learned inference step that moves the mesh nodes. This is why we can separate these two, monitor preprocessing and mesh movement, in the timings in Table\,\ref{mv_benchmark}. In particular we demonstrate that above the time savings in the overall mesh movement process, there a significant speedup, by a factor of 40, of the core mesh movement can be achieved by running UM2N on GPU. It should be noted that we expect, in future work, that further improvements can be made in reducing the monitor-preprocessing stage costs, e.g. by switching to an Arbitrary Langrian-Eulerian formulation for interpolation, thus further improving the overall runtime reduction offered by UM2N.

These results demonstrate that UM2N method provides a measurable improvement not only in resolving the region of tsunami inundation along the shoreline but also in tracking the free-surface elevation at gauge locations, while offering a favorable balance between accuracy and computational cost.
\begin{table}[t]
  \centering
  \resizebox{.85\textwidth}{!}{
  \begin{tabular}{lrrr}
    \toprule
    \midrule
    Mesh configuration
      & \makecell{$\text{RMSE}_{\text{run-up}}$ ($\unit{cm}$)}
      & \makecell{$\text{ER}_{\text{run-up}}$ (\%) $\downarrow$}
      & \makecell{$\text{ER}_{\text{Gauge 3}}$ (\%) $\downarrow$}
      \\
    \midrule
    FIX-COARSE
      & $1.911$
      & --
      & --
      \\
    ADAPT-UM2N
      & $1.303$
      & $31.82$
      & $\bf{91.14}$
      \\
    ADAPT-MA
      & $1.057$
      & $\bf{44.69}$
      & $73.64$
      \\
    \midrule
    \bottomrule
  \end{tabular}
  }
  \caption{\label{tab:maxrunup_mv} Quantitative results across the fixed coarse mesh, UM2N adapted mesh and MA adapted mesh for Monai Valley test case. $\text{RMSE}_{\text{run-up}}$: root mean square (RMS) error of the maximum run-up position relative to the fixed fine mesh (reference solution), measured as the difference in the furthest inland $x$-coordinate reached by the wetting-drying boundary at each $y$-position along the shoreline. $\text{ER}_{\text{run-up}}$: RMS error reduction of maximum run-up relative to the fixed coarse mesh. $\text{ER}_{\text{Gauge 3}}$: Error reduction of the wave peak elevation at Gauge 3 relative to the fixed coarse mesh (see Fig.\,\ref{fig:elev_mv}).}
\end{table}

\begin{table}[t!]
  \centering
  \resizebox{1.\textwidth}{!}{
  \begin{tabular}{lrrr}
    \toprule
    \midrule
      \text{ }
      & \makecell{MA}
      & \makecell{UM2N}
      & \makecell{UM2N} \\
    Hardware
      & \makecell{CPU}
      & \makecell{CPU}
      & \makecell{GPU (Nvidia A100)} \\
    \midrule
    Total runtime breakdown (s)
      &  $18807$
      &  $14033$
      &  $12821$\\
    Total mesh movement (s)/fraction (\%)
      &  $12563$ ($66.80$)
      &  $7760$ ($55.30$)
      &  $5999$ ($46.79$) \\
    Mesh inference (s)/fraction (\%)
      &  --
      &  $1748$ ($12.46$)
      &  $43$ ($0.34$) \\
    \midrule
    Total runtime reduction (\%) $\downarrow$
      & -- 
      & $25.39$
      & $\bf{31.83}$\\
    Total mesh movement speedup $\uparrow$
      & -- 
      & $\sim 1.62\times$
      & $\bf{\sim 2.09\times}$\\
    UM2N inference speedup $\uparrow$
      & -- 
      & --
      & $\bf{\sim 40\times}$\\
    \midrule
    \bottomrule
    \end{tabular}
    }
  \caption{\label{mv_benchmark} Performance benchmark on the Monai Valley test case. Average total runtime and mesh movement time are recorded over five runs for both UM2N and MA movement. ``Total mesh movement'' time contains the complete mesh adaptation pipeline: monitor function computation, monitor smoothing, and the core mesh movement step. ``Mesh inference'' time refers to the core mesh movement step only (UM2N forward pass). Runtime reduction is computed as the percentage decrease relative to the MA mesh configuration. Total mesh movement speedup is the ratio of MA total mesh movement time on CPU to that of each UM2N mesh configuration. UM2N inference speedup is the ratio of UM2N  core mesh movement inference time on CPU to that on GPU.}
\end{table}

\section{Conclusion}
\label{conclusion}
This paper demonstrates the integration of the UM2N machine learning based mesh movement method with a depth-integrated non-hydrostatic shallow water model within the Thetis/Firedrake framework, applied to coastal tsunami simulation with dispersive wave effects.

Four test cases of increasing complexity are used to evaluate our approach. The major findings are summarized as follows: (1) In the N-wave strip source test case, both UM2N and MA mesh movement achieve lower RMS errors than fixed meshes across all mesh resolutions, with UM2N providing a $\sim 2\times$ speed-up over conventional MA movement in total mesh movement step (see Fig.\,\ref{fig:N_wave_convergence}). (2) In the test cases of solitary wave over a truncated conical shoal and conical island, the MA solver diverges within $70$--$100$ time steps due to non-convergence of the iterative Monge--Amp\`{e}re equation solver, caused by the steep monitor function gradients arising from wave refraction and wetting-drying interfaces. Compared with MA movement, UM2N maintains stable mesh adaptation throughout the full simulation period in both scenarios, demonstrating its superior robustness under strongly nonlinear wave conditions. For the conical island case, spatial analysis of shoreline positions and maximum run-up outlines further shows that UM2N achieves a $\sim34\,\%$ reduction in the RMS error of the maximum run-up position relative to the fixed coarse mesh. (3) In the Monai Valley laboratory benchmark, the maximum run-up analysis along the shoreline indicates that UM2N achieves a $\sim32\,\%$ reduction in the RMS error of the run-up position, which is comparable to the $\sim45\,\%$ reduction obtained by the MA movement. Additionally, UM2N achieves a $\sim91\,\%$ reduction in wave-peak error at Gauge 3 relative to the fixed coarse mesh, compared with $\sim74\,\%$ for the MA method, while reducing total runtime by $\sim32\,\%$ and accelerating the total mesh adaptation inference by $\sim 2$ times on GPU (see Tables\,\ref{tab:maxrunup_mv}--\ref{mv_benchmark}). (4) New monitor functions integrated with a wetting-drying interface tracker are designed and shown to correctly guide mesh movement in scenarios involving complex shoreline dynamics.

These results indicate that UM2N is an effective surrogate for the costly MA mesh movement, providing comparable run-up accuracy while significantly reducing computational cost. They also demonstrate that the tight coupling of machine learning based mesh movement with a DG-FE based non-hydrostatic PDE solver is feasible and practical for coastal tsunami applications.

The UM2N based PDE solver framework requires repeated solution transfer between adapted meshes at each timestep. In the present work, this transfer is performed via the supermeshing based Galerkin projection based on the work of \cite{Farrell2009} and \cite{Farrell2011}, which preserves the global integral of the projected field. The same remapping procedure is adopted for both MA and UM2N mesh configurations. Although a formal quantitative assessment of conservation errors introduced by repeated remapping is beyond the scope of this study, the long-time agreement of the UM2N solutions with analytical solutions, fine mesh reference computations and laboratory measurements provides practical evidence that the repeated mesh-to-mesh transfer does not introduce significant cumulative error in the reported simulations.

Several limitations of our current work motivate directions for future research: (1) The UM2N model used in this study was trained on two-dimensional unit meshes with $463$--$513$ vertices, while the test cases employed meshes with up to $\sim10^4$ degrees of freedom. Building upon the work of \cite{LI20054915} and \cite{Rowbottom2025}, extending the framework to three-dimensional (3D) mesh geometries would broaden the applicability to full 3D ocean simulations. (2) Although UM2N demonstrates strong robustness where MA method fails, the UM2N adapted meshes display somewhat less regular element shapes and sizes away from the wave features compared with MA movement (see Fig.~\ref{fig:mesh_mv}). As discussed in Sects.\,\ref{N-wave} and \ref{monaivalley_testcase}, UM2N tends to concentrate mesh refinement very locally around steep free-surface gradients, resulting in relatively irregular element sizes in surrounding regions. Therefore, mesh adaptation algorithms could be further improved by introducing constraints that explicitly control element quality during mesh movement (e.g., enforcing smoother transitions in element size between refined and coarse regions). Additionally, suggested by~\cite{zhang2024}, allowing boundary node movement would also help generate meshes with superior regularity. (3) Unsupervised learning approaches such as G-adaptivity~\citep{Rowbottom2025}, and Generalizable Mesh Movement Network (UGM2N)~\citep{Wang2025UGM2N} provide alternative movement strategies that eliminate the need for MA-generated training data, which potentially improves generalization to new computational domains and new monitor functions beyond those seen during training.

\section*{Acknowledgements}
YL would like to acknowledge Mingrui Zhang and Chunyang Wang for their invaluable advice and technical support on the use of the UM2N model. YL also gratefully thanks Wei Pan for helpful advice in benchmarking setup. All authors would like to acknowledge the Imperial College London Research Computing Service for their support.

\section*{Software and Data Availability}
The source code for the UM2N based non-hydrostatic model presented in this work is available from \href{https://github.com/yezhang-li4222/UM2N_NH}{GitHub Repository}.











\bibliographystyle{elsarticle-harv}
\bibliography{bibliography_rev}
\end{document}